\documentclass[fleqn,usenatbib]{mnras}
\usepackage{newtxtext,newtxmath}

\usepackage[T1]{fontenc}

\DeclareRobustCommand{\VAN}[3]{#2}
\let\VANthebibliography\thebibliography
\def\thebibliography{\DeclareRobustCommand{\VAN}[3]{##3}\VANthebibliography}

\usepackage{graphicx}	
\usepackage{amsmath}	
\usepackage{xspace}

\RequirePackage{lineno}
\usepackage[utf8]{inputenc}
\usepackage{ae,aecompl}
\usepackage{comment}
\DeclareUnicodeCharacter{2212}{-}
\usepackage{natbib}
\usepackage{float}
\usepackage{graphicx}
\usepackage{xcolor}
\usepackage{rotating,times,pictex,graphicx,latexsym}
\usepackage{color}
\usepackage{amsmath}
\usepackage{threeparttable}
\usepackage{lipsum}
\usepackage{booktabs}
\usepackage[all]{hypcap}
\usepackage[title,titletoc]{appendix}
\usepackage[]{hyperref}
\PassOptionsToPackage{pdfpagelabels=false}{hyperref}
\usepackage[T1]{fontenc}

\newcommand{\Ms}{M\ensuremath{_{\odot}}}

\newcommand{\beq}{\begin{equation}}
\newcommand{\eeq}{\end{equation}}

\newcommand{\mstar}{\ensuremath{M_\ast}}

\newcommand{\hii}{\ion{H}{ii}\xspace}

\newcommand{\arcs}{\hbox{$^{\prime\prime}$}}

\newcommand{\arcm}{\mbox{$^{\prime}$}}

\newcommand{\into}{\mbox{$\times$~}}
\newcommand{\lbol}{\mbox{\rm L$_{\rm bol}$}~}

\newcommand{\lsun}{\mbox{\rm L$_{\odot}$}}

\newcommand{\degree}{\mbox{$^{\circ}$}}
\newcommand{\av}{\mbox{A$_\mathrm{V}$~}}
\newcommand{\ak}{\mbox{A$_\mathrm{K}$~}}

\newcommand{\kms}{\hbox{km~s$^{-1}$}}
\newcommand{\cms}{\hbox{cm$^{-2}$~}}

\newcommand{\cmq}{\hbox{cm$^{-3}$}}

\newcommand{\vlsr}{\hbox{V$_{\rm LSR}$}}

\newcommand{\nht}{\hbox{N$(\mathrm H_2)$}}

\newcommand{\thco}{\hbox{$^{13}$CO$_{1-0}$}~}
\newcommand{\twco}{\hbox{$^{12}$CO}~}

\newcommand{\cloud}{{\rm G148.24+00.41}}

\title[Cluster Formation in GMC G148.24+00.41]
{
Probing the Global Dust Properties and Cluster Formation Potential of
the Giant Molecular Cloud G148.24$+$00.41
}

\author[Rawat et al.]{
Vineet Rawat,$^{1,2}$\thanks{E-mail: vineet@prl.res.in}
M. R. Samal,$^{1}$
D. L. Walker,$^{3}$
A. Zavagno,$^{4,5}$
A. Tej,$^{6}$
G. Marton,$^{7}$
D.K. Ojha,$^{8}$\newauthor
Davide Elia,$^{9}$
W.P. Chen,$^{10}$
J. Jose,$^{11}$
C Eswaraiah,$^{11}$
\\
$^{1}$Physical Research Laboratory, Navrangpura, Ahmedabad, Gujarat 380009, India\\
$^{2}$Indian Institute of Technology Gandhinagar Palaj, Gandhinagar 382355, India\\
$^{3}$Jodrell Bank Centre for Astrophysics, Department of Physics and Astronomy, University of Manchester, Oxford Road, Manchester M13 9PL,UK\\
$^{4}$Aix-Marseille Universite, CNRS, CNES, LAM, 38 rue F. Joliot Curie, 13388 Marseille Cedex 13, France\\
$^{5}$Institut Universitaire de France, Paris, 1 rue Descartes, 75231 Paris Cedex 05, France\\
$^{6}$Indian Institute of Space Science and Technology (IIST), Thiruvananthapuram 695 547, Kerala, India\\
$^{7}$Konkoly Observatory, Research Centre for Astronomy and Earth Sciences,
Hungarian Academy of Sciences, H-1121 Budapest,\\
$^{8}$Department  of  Astronomy  and  Astrophysics,  Tata  Institute  of  Fundamental  Research,  Mumbai  400005, India\\
$^{9}$Istituto di Astrofisica e Planetologia Spaziali, INAF, Via Fosso del Cavaliere 100, I-00133 Roma, Italy\\
$^{10}$Institute of Astronomy, National Central University, Jhongli 32001, Taiwan\\
$^{11}$Indian Institute of Science Education and Research (IISER) Tirupati, Rami Reddy Nagar, Karakambadi Road,
Tirupati 517 507, India\\
}

\begin{document}
\label{firstpage}
\pagerange{\pageref{firstpage}--\pageref{lastpage}}
\maketitle


\begin{abstract}
Clouds more massive than about $10^5$~\Ms~are potential sites of massive cluster formation. Studying the properties of such clouds in the early stages of their evolution offers an opportunity to test various cluster formation processes.
We make use of CO,  $\it{Herschel}$, and UKIDSS observations to study one such cloud, \cloud.
Our results show the cloud to be of high mass ($\sim$ $1.1\times10^5$ \Ms), low dust temperature ($\sim$ 14.5\,K), nearly circular (projected radius $\sim$ 26 pc), and gravitationally bound with a dense gas fraction of $\sim 18$\% and a density profile with a power-law index of $\sim -1.5$. Comparing its properties with those of nearby molecular clouds, we find that \cloud~is comparable to the Orion-A molecular cloud in terms of mass, size, and dense gas fraction.
From our analyses, we find that the central area of the cloud is actively forming protostars  and is moderately fractal with a Q-value of $\sim$ 0.66. We also find evidence of global mass-segregation in the cloud, with a degree of mass-segregation ($\Lambda_{MSR}) \approx3.2$. We discuss these results along with the structure and compactness of the cloud, the spatial and temporal distribution of embedded stellar population, and their correlation with the cold dust distribution, in the context of high-mass cluster formation. Comparing our results with models of star cluster formation, we conclude that the cloud has the potential to form a cluster in the mass range $\sim$ 2000--3000 \Ms~through dynamical hierarchical collapse and assembly of both gas and stars.

\end{abstract}




\section{Introduction}
\label{int}

Giant molecular clouds (GMCs) are the cradles of young star clusters \citep{che22} in which most stars form \citep{lada03}. Massive to intermediate-mass clusters play a dominant role in the evolution and chemical enrichment of the Galaxy via stellar feedback such as photoionization, stellar winds, and supernovae \citep[e.g.][]{green15, green16, kim18}. Understanding cluster formation, in particular the formation of intermediate-mass ($10^3 -10^4 $ \Ms; \citealp{wei_15}) to high-mass clusters ($>10^4$ \Ms; \citealp{por10}), is therefore crucial and one of the key problems in modern astrophysics \citep[e.g.][]{long14,kra20, krum20}. Young massive clusters \cite[YMCs;][]{por10} are thought to be the potential modern-day analogues of globular clusters (GCs) that formed in the early Universe. Determining their formation mechanism will better constrain the different cluster formation models over the full mass range of clusters.

The general consensus is that GMCs convert $\sim 3$--10\% of their mass into stars before being dispersed \citep{evans09, lada2010}. In this regard, massive bound clouds with mass $\geq 2\, \times 10^5$ \Ms~are the potential formation sites for massive stellar clusters of mass $> 10^4$ \Ms, assuming the star-formation efficiency is as low as 5\%. Only a handful of YMCs have been found in our Galaxy \citep[e.g. see Table 2 of][]{por10}, despite the fact that it hosts many high-mass clouds \citep[e.g. see Figure 2 of][]{sol87}. 

Studies of Galactic disk clouds, suggest that clouds with a relatively high dense gas 
fraction \citep[i.e. fraction of gas with $n$ $\geq$ 10$^4$ \cmq, or N(H$_2$) $\geq$ 6.7 $\times$ 10$^{21}$ \cms  with respect to the total gas of the cloud;][]{lada2010} are the sites of richer star formation \citep{lada12,evans14}, while other
studies suggest that the Galactic environment plays a significant role in defining the initial conditions of star-formation in molecular clouds \citep[e.g. Galactic centre clouds, see review by][]{Hen_2022}. The geometry and structure of molecular clouds also likely play a crucial role in the formation and growth
of star clusters \citep[e.g.][] {Burk-Har2004,heitsch08, Pon2012, clark2015, heigl2022, Hoe2022}.


Determining and evaluating the physical conditions, kinematics, structures, and dynamics of GMCs at their initial stages of evolution are crucial for finding the favourable conditions that a GMC requires to produce a massive cluster.
In this context, massive clouds of mass  $\geq 10^5$ \Ms, where stellar feedback effects are not yet significant, are potential targets for studying early phases of cluster formation. The molecular cloud - \cloud, the focus of the present work, is one such cloud whose properties, structure, and star-formation potential remain largely unexplored.

\subsection{G148.24+00.41 Cloud}
\label{into}

At submillimeter and millimeter wavelengths, Planck observations \citep{pla11a} have revealed a population of dense molecular cold clumps and clouds in our Galaxy. The Planck Early Release Compact Source Catalog  \citep[PERCSC;][]{pla11} provides lists of positions and flux densities of the sources in nine frequencies in the range 30--870 GHz with beam sizes ranging from 33\arcm~to 5\arcm.  The PERCSC source ``G148.24+00.41'' corresponds to a  cloud (ID: 2182) identified by
\citet{mivi17}, using low-resolution (beam $\sim$ 8.5\arcm) $^{12}$CO data at 115 GHz. 
Its temperature as estimated by Planck is around 13.5 K, while its mass  as estimated by \citet{mivi17} is $\sim$ 1.3 \into 10$^{5}$ \Ms,
suggesting that G148.24+00.41 is a massive cold cloud. The cloud area
also includes the dark cloud ``TGU 942P7'', identified by \citet{dob05}
based on the digitized sky survey extinction map. The average visual extinction (A$\mathrm{_V}$) derived in the direction of  TGU 942P7 using 2MASS data  is around $\sim$ 5.1 mag \citep{dob11}.
The Infrared Astronomical Satellite (IRAS) also identified a source ``IRAS 03523+5343'' in the direction of \cloud~close to TGU 942P7. 

\begin{figure}
    \centering
    \includegraphics[width=8.5cm]{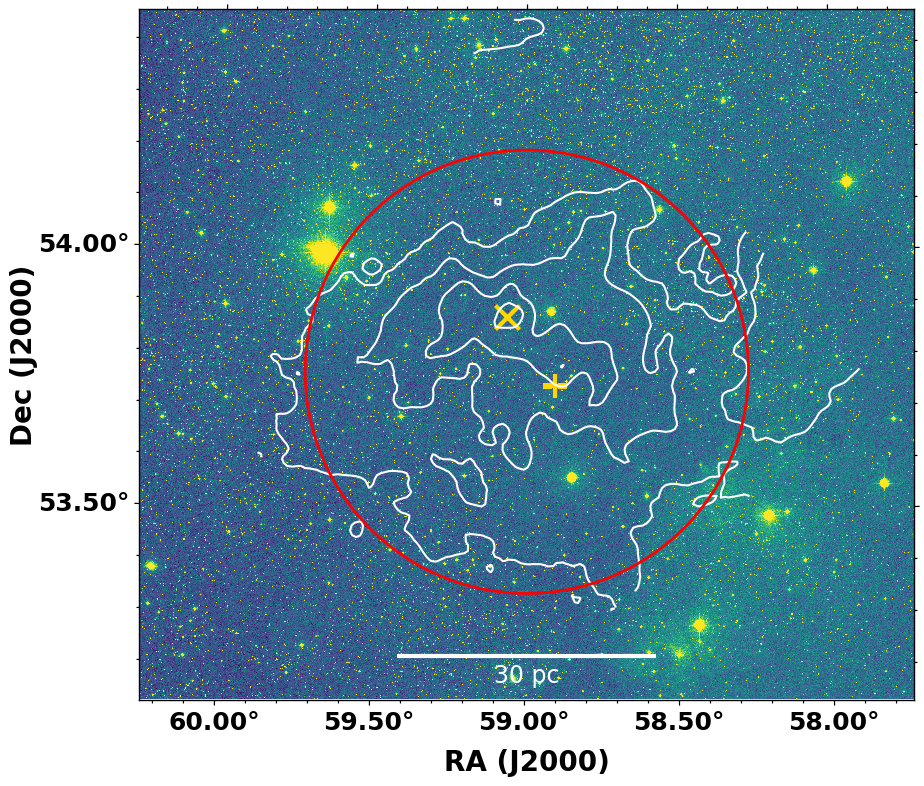}
    \caption{DSS2 R-band optical image of the \cloud~cloud for an area of $\sim$ 1.9\degree $\times$ 1.3\degree overlaid with the contours of \twco(J = 1$-$0) emission (discussed in Section \ref{dis}), integrated in the velocity range $-$37 to $-$30 \kms. The contour levels are at  1.5, 10, 20, 30, and 40.0 K\, \kms. The red solid circle (centered at: $\alpha$ = 03:55:59.02 and $\delta$ = +53:45:48.03) shows the overall extent of the cloud of radius $\sim$ 26 pc.
    The plus and cross sign represents the position of TGU 942P7 and IRAS 03523+5343, respectively.}
    \label{fig_dss2}
\end{figure}

Figure \ref{fig_dss2} shows the optical view of the cloud
along with the locations of TGU 942P7 and IRAS 03523+5343. The peak velocity of the various molecular gas associated with the IRAS 03523+5343
source and its immediate vicinity, estimated by different authors, lies mostly in the range $\sim$ $-$33 to $-$35 \kms~\citep[e.g.][]{wb1989,yang02,urqh_2008, mivi17}.
The kinematic distance of the cloud as found in the literature lies in the range 3.2--4.5 kpc \citep{yang02,coop13,maud15, mivi17}. In
the direction of the cloud, the signature of star formation in terms of young stellar objects (YSOs) \citep{wins2020} and
cold cores \citep{yuan16, zhag18} have been identified.

Despite the fact that \cloud~is a  cold massive cloud, 
its global properties, structure, physical conditions, and stellar content have not been studied in detail. In this work, we explore these components of \cloud~with the aim to understand its cluster formation potential, and to constrain the mechanism(s) by which an eventual YMC may form.

We organize this paper as follows. In Section \ref{data}, we present an overview of the data sets used. In Section  \ref{res}, we
present the measured global dust properties and physical conditions of \cloud, and compare the results with nearby clouds. We discuss its  protostellar contents and their spatial and luminosity
 distributions. We also examine the stellar clustering structure and evidence of mass-segregation. In Section
 \ref{cfm}, we discuss the cluster
 formation processes and potential for the cloud to form a high-mass cluster. We summarize our findings
 in Section \ref{s28_conc}.

\section{Data}
\label{data}
 We used near-IR ($J, H,$ and $K$) photometric catalogues from the UKIDSS 10th data release \citep{lawrence07}. These catalogues are the data products of the UKIDSS Galactic Plane Survey  \citep[GPS;][]{lucas2008}, done using the observations taken with the UKIRT 3.8-m
telescope. The UKIDSS GPS data has saturation
limits at J $=$ 13.25, H $=$ 12.75 and K $=$ 12.0 mag \citep{lucas2008}. For sources brighter than these above limits, we have used 2MASS photometry values. We consider only those sources that have photometric uncertainty $<$ 0.2 mag in all three bands for our analysis. The GPS data are $\sim$ 3 magnitude deeper than 2MASS data, thus, would give a better assessment of
extinction than those measured by \citet{dob11}.

We used far-infrared images of the Herschel Infrared Galactic Plane Survey \citep[Hi-GAL;][]{Mol_2010a}, taken with the $Herschel$ PACS and SPIRE instruments, centred on wavelengths of 70, 160, 250, 350, and 500 $\micron$. The angular resolution of these images are 8.5, 13.5, 18.2, 24.9, and 36.3 arcsec, respectively \citep{Mol_2010a}. 

We also used \twco(J = 1--0) emission molecular data at 115 GHz,  observed with the 13.7-m radio
telescope as a part of the Milky Way Imaging Scroll Painting survey \citep[MWISP;][]{su19}, led by the Purple Mountain Observatory (PMO). The angular resolution of the CO data is $\sim$ 50\arcsec (or 0.8 pc at the distance of 3.4 kpc; see Sect. \ref{dis} for distance), while its spectral resolution is $\sim$ 0.16 \kms. The typical sensitivity per spectral channel is about 0.5 K \citep[for details, see][]{su19}. 
This data brings a factor of $\sim$ 10 improvement in the spatial resolution and a factor of 8 in the velocity resolution compared to the previous CO survey data \citep[beam $\sim$ 8.5\arcm, velocity resolution $\sim$ 1.3 \kms;][]{Dame2001} used by \citet{mivi17} to identify the cloud.

\begin{figure}
    \centering
    \includegraphics[width=8cm]{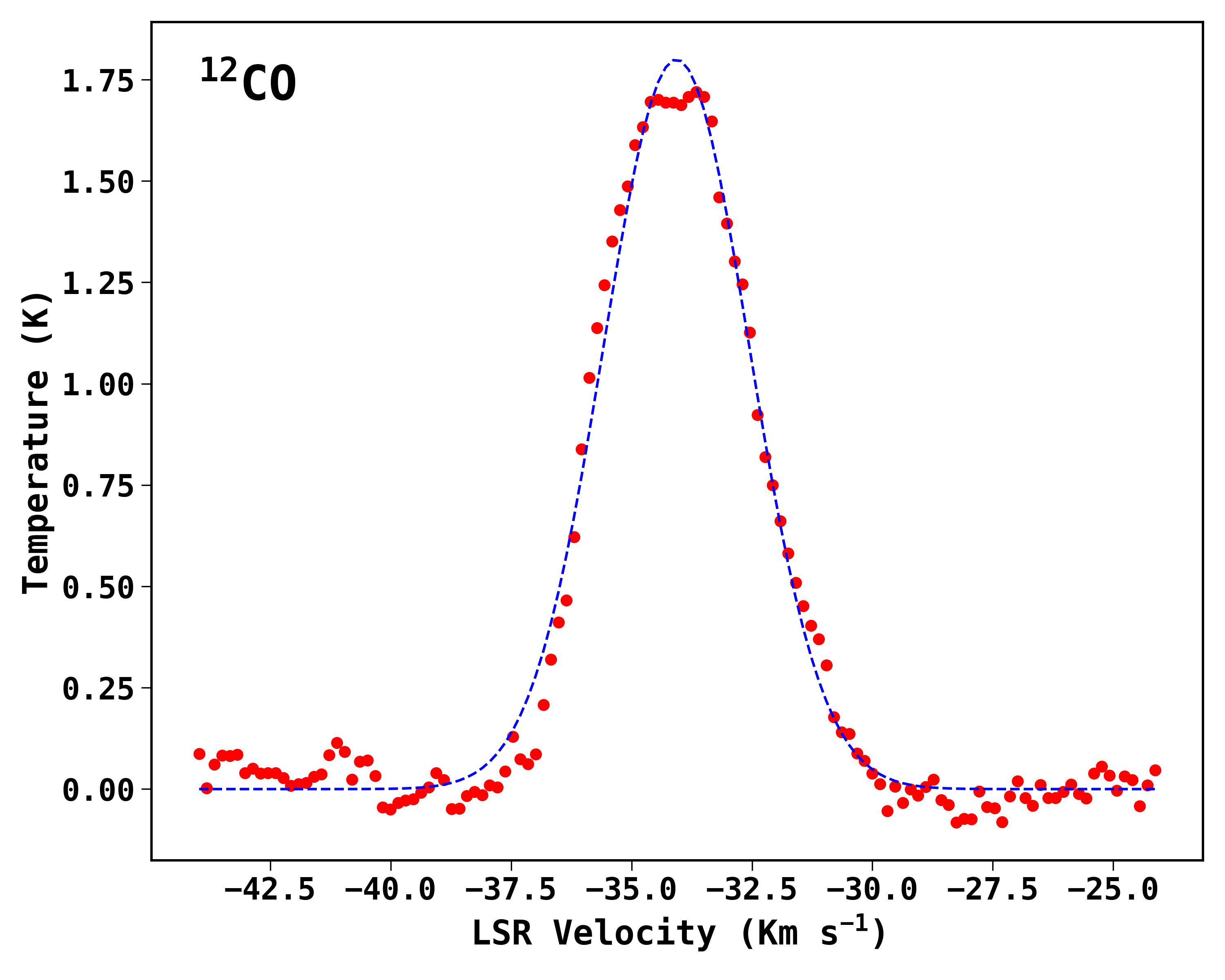}
    \caption{The average $\mathrm{^{12}CO}$ spectral profile towards the direction of \cloud. The dashed blue line represents the fitted Gaussian profile. }
    \label{fig_co}
\end{figure}

\section{Analysis and Results}
\label{res}
\begin{figure*}
    \centering
    \includegraphics[width=8.8cm]{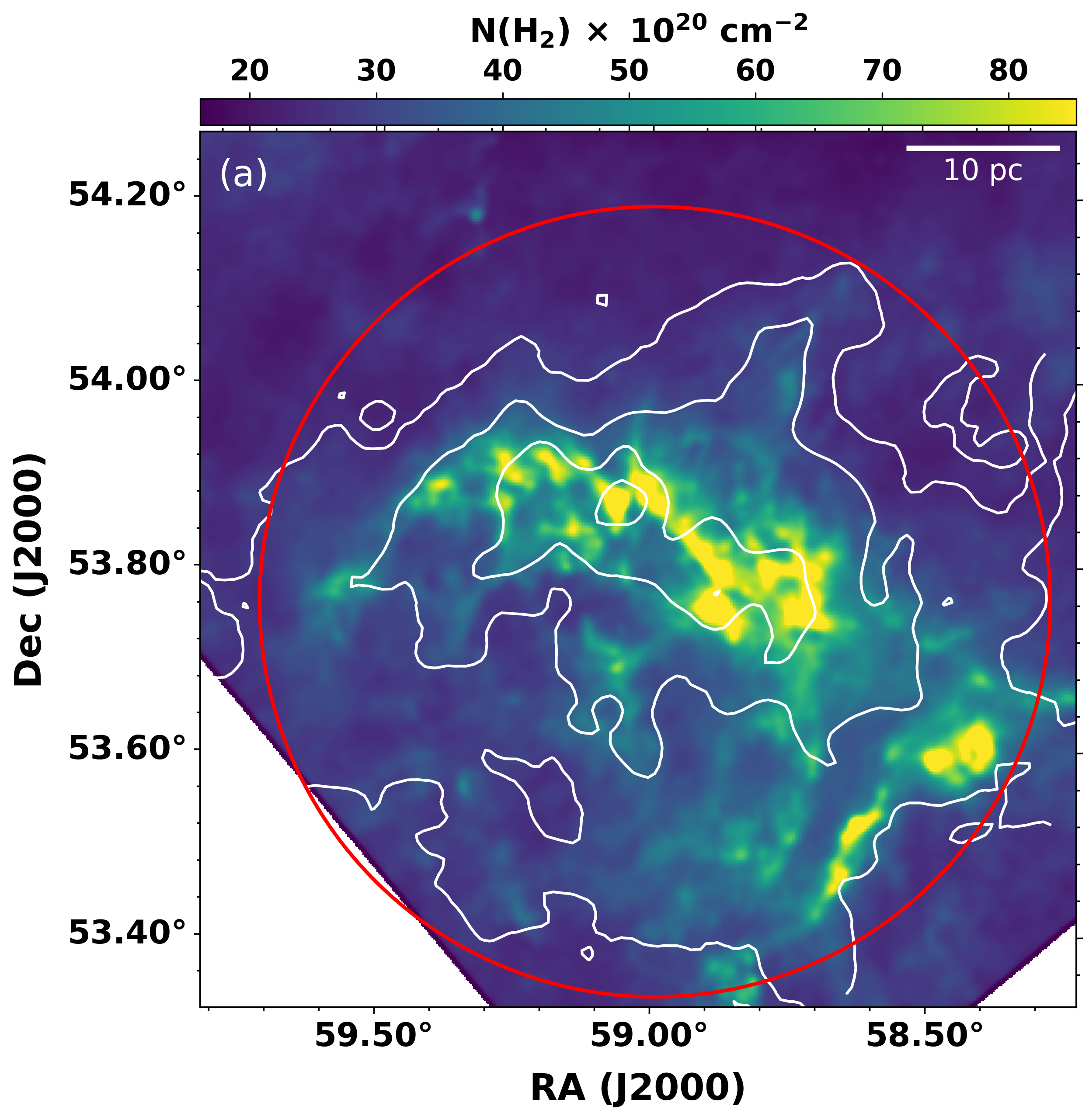}
    \includegraphics[width=8.9cm, height=8.97cm]{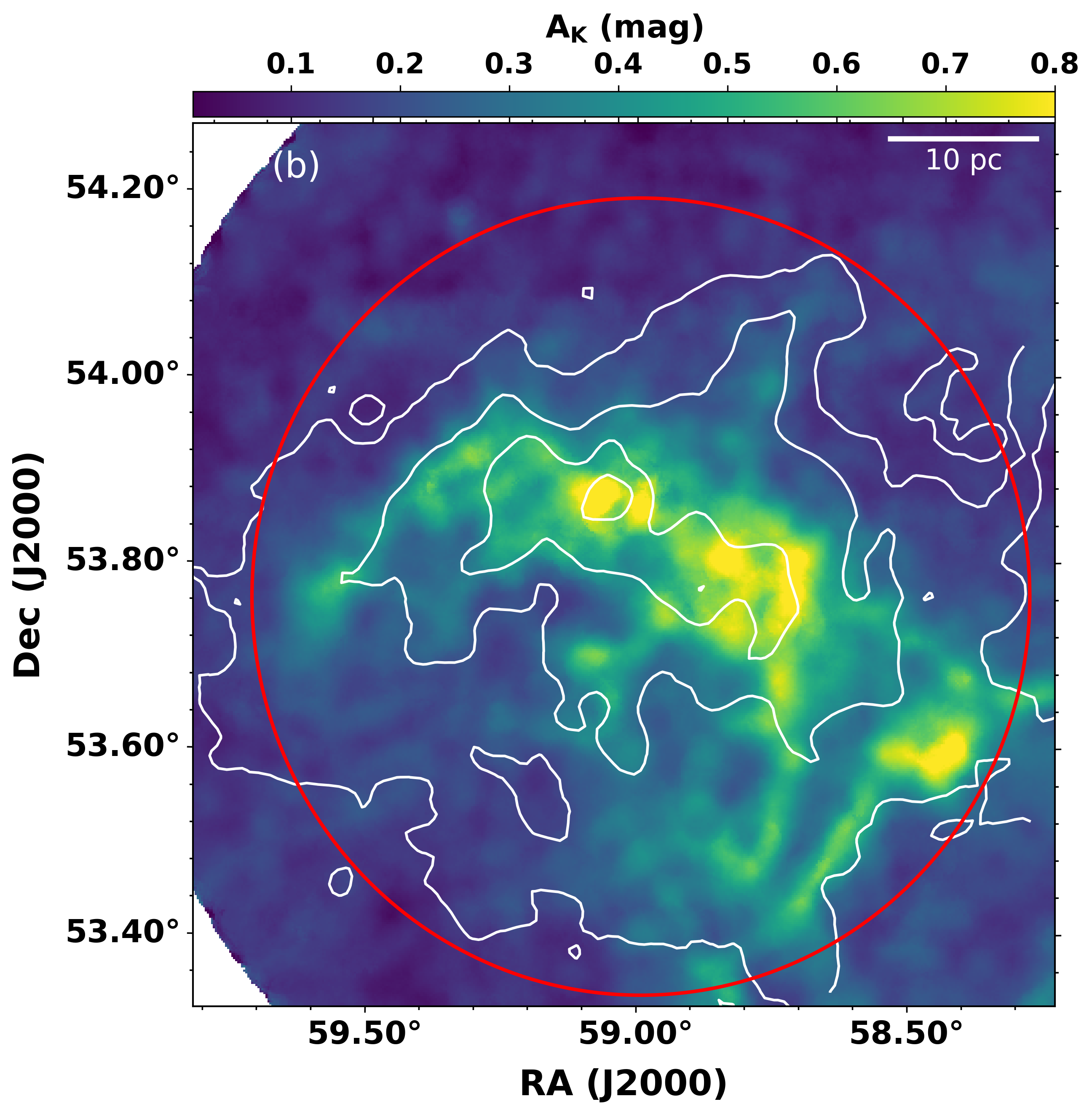}
    \caption{(a) $\it{Herschel}$ column density map (resolution $\sim 12\arcs$) and (b) K-band extinction map (resolution $\sim 24\arcs$), over which the contours of CO integrated emission are shown. The contour levels are the same as in Figure \ref{fig_dss2}. The solid red circle denotes the boundary of the cloud.}
    \label{fig_dust}
\end{figure*}

\begin{figure}
    \centering
    \includegraphics[width=8.7cm]{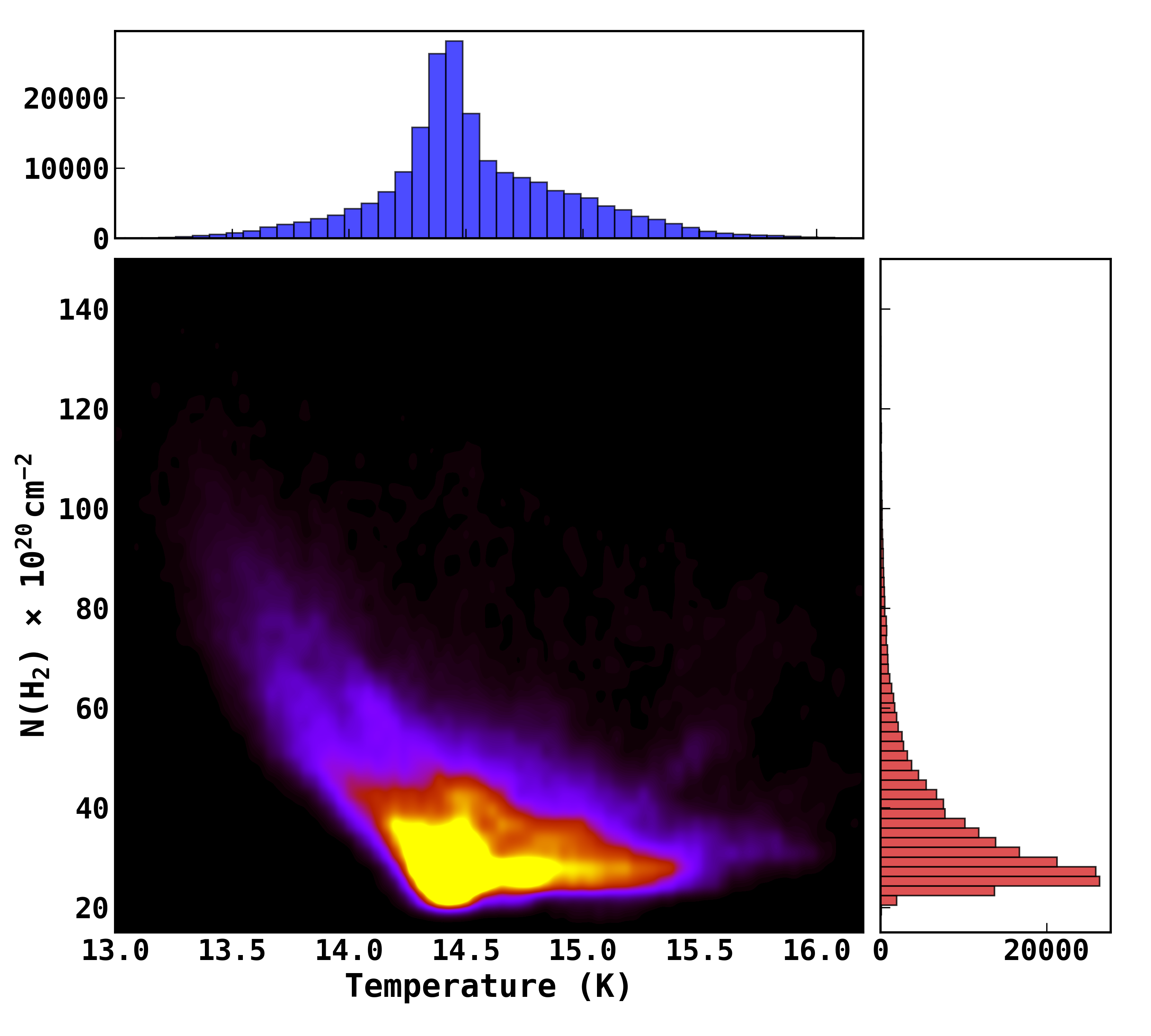}
    \caption{Column density versus temperature diagram, showing the distribution of physical  conditions of dust in \cloud. 
    }
    \label{fig_phy}
\end{figure}

\subsection{Distance, Physical Extent, and Large-Scale Gas Morphology}
\label{dis}
Figure \ref{fig_dss2} shows the DSS2 R-band optical image of the \cloud~cloud along with the contours  of the $^{12}$CO intensity emission, integrated in the velocity range $-$37 to $-$30 \kms. 
We determine the peak  systemic velocity of the cloud with respect
to the local standard of rest ($\vlsr$), the 1D velocity
dispersion ($\sigma_{1d}$), and the associated velocity range as $-$34.07 $\pm$ 0.02 \kms, 1.51 $\pm$ 0.02 \kms, and [$-$37, $-$30] \kms, respectively. This is done by fitting a Gaussian profile to the 
CO spectrum of \cloud~and is shown in Figure \ref{fig_co}. We estimated the line-width
($\Delta V = 2.35\, \sigma_{1d}$) and 3D velocity dispersion ($\sigma_{3d} =  \sqrt{3} \times \sigma_{1d}$) associated with the
CO profile as 3.55 and 2.62 \kms, respectively. We note that the CO emission shows a flattened shape around the peak. However, such a flattened profile is not found in the \thco~spectrum (velocity resolution $\sim$ 0.14 \kms) of the source, observed by  \cite{urqh_2008}. This suggests that probably due to the high optical depth, self-absorption occurs in the $^{12}$CO line, resulting in the flattened top seen in the line-profile.

The kinematic distance of the cloud is quite uncertain (discussed in Section \ref{int}). We thus recalculated the distance using the Monte Carlo-based kinematic distance calculation code\footnote{https://github.com/tvwenger/kd} \cite[described in][] {weng18}, $\vlsr$ value as  $-34.07 \pm 0.02\, \kms$, and considering the recent galactic rotation curve model of \cite{reid_2019}. 
We ran the simulation 500 times and found the resultant kinematic distance to be $\sim$ 3.4 $\pm$ 0.3 kpc, which is used in this work. 
We note, since  the \cloud~cloud is located in the outer galaxy, no near-far kinematic distance ambiguity is present for the cloud.

As can be seen from Figure \ref{fig_dss2}, the cloud in its central area shows a non-uniform and elongated intensity distribution of CO gas, but overall in CO, it appears to be a nearly circular structure of radius $\sim$ 26 pc on the plane of the sky.
The radius bordering the outer extent of the cloud is marked in the figure by a red circle.
From Figure \ref{fig_dss2}, one may also notice that the cloud is devoid of optically visible star clusters.
The cloud is also found to be devoid of \hii regions when searched in the H$_\alpha$  survey images  of the Northern Galactic Plane \citep{bar14} and the 6 cm radio continuum images of the Red MSX Source (RMS) survey \citep{lums13}. 
The non-detection of such sources implies that the cloud is in its early phases of evolution and strong stellar feedback is yet to commence in the cloud.

\subsection{Global Dust Properties and Comparison with  nearby GMCs}
\label{gdp}
Molecular clouds are characterized by a gas column density corresponding to visual extinction, \av $\geq$ 1--2 magnitudes. For example, \citet{lada2010}, using near-infrared extinction maps, derived cloud masses of a number of nearby ($<$ 500 pc) molecular clouds, including GMCs like Orion-A, Orion-B, and California, by integrating cloud area above K-band extinction, \ak $\geq$ 0.1 mag.
Similarly,  for several nearby molecular clouds, \citet{hei10} estimated cloud masses by integrating cloud area above visual extinction, \av $>$ 2 mag. We note that \av $=$ 2 magnitude corresponds to \ak $\sim$ 0.2 mag using the relation, \ak = 0.112 $\into$ \av, from \citet{rie85}.

In nearby molecular clouds, it has been found
that young stars that are formed above an extinction threshold  
of  \ak $\geq$ 0.8 mag (or  equivalent column density  $\geq$
6.7 \into 10$^{21}$ \cms)  are well correlated with the corresponding
gas mass \citep{lada2010, hei10}. In fact, \citet{lada12} find that above this extinction threshold, a linear relationship between the star-formation rate 
and the column density is clearly apparent. \citet{lada2010, lada12} advocated that since above \av$>$ 6 mag, dense gas tracer molecules such as HCN and N$\mathrm{_2H^{+}}$ have been observed in molecular clouds, thus, a column density above \ak$>$ 0.8 (or \av >  7 mag) mag represents the dense gas content of the molecular clouds.
Following the same convention, we  also use this threshold to estimate the dense gas properties of the \cloud~cloud.

We estimated the total (\ak$>$ 0.2 mag) and the dense gas (\ak $>$ 0.8 mag) properties of \cloud~in two ways: i) using the $Herschel$ column density map, and ii) using the UKIDSS based near-infrared extinction map. The latter is mainly used to compare the global properties of \cloud~with the properties of the nearby GMCs studied by \cite{lada2010} and \citet{hei10}.

\subsubsection{Properties from Dust Continuum Map}
\label{dust}
\cite{marsh17} constructed the dust temperature (T$_{\rm d}$) and the column density (\nht) maps
of the inner Galaxy using $Herschel$ data, collected as a part of the Hi-GAL Survey \citep{Mol_2010a}. They constructed maps using the PPMAP technique \citep[see][for details]{mar15}, resulting in high-resolution ($\sim$ 12\arcs) dust temperature and column density maps. PPMAP technique considers the point spread functions (PSFs) of the telescopes that enable to use the images at their native resolution, and also drops the assumption of uniform dust temperature 
along the line of sight. Thus, the PPMAP data represents a significant improvement over those obtained with a more conventional spectral energy distribution (SED) fitting technique, in which a pixel-to-pixel modified black-body fit to the Herschel images is done after convolving them to the resolution of the 500 $\mu$m band. While fitting, the dust temperature is assumed to be uniform everywhere along the line of sight and the dust opacity index is often assumed as 2 \citep[e.g.][]{Batt_2011, deh12, Konyves_2015, schi2020}. 
Owing to better resolution as well as its ability to account for the line-of-sight temperature variation, PPMAP data have been used in the analysis of several molecular clouds \cite[e.g.][]{mar19, spi21}.

Figure \ref{fig_dust}a shows the $Herschel$ column density map overlaid with the contours of CO emission.
As can be seen, the morphology of the column density map correlates well with the overall CO emission, particularly in the central area.
Owing to high-resolution,
the column density map in the central area of the cloud shows more clumpy and filamentary structures. Figure \ref{fig_phy} shows the \nht~versus T$_{\rm d}$ distribution within the cloud boundary, showing that they are inversely correlated as seen in infrared dark clouds \citep[e.g.][] {Batt_2011}.
Within the cloud boundary, we find that the column density  lies in the range 2.0--40.0 $\times$ 10$^{21}$ cm$^{-2}$  with a median value of $\sim$ 3.2 $\times$ 10$^{21}$ cm$^{-2}$, while
the dust temperature lies in the range 12.7--21.3 K, with a median value of $\sim$ 14.5 K. 

We measured the global properties  of the cloud by considering all the pixels within the cloud area whose \nht~value is greater than $20 \times 10^{20}~ \cms$. Using the empirical relation, \av = N$(\mathrm{H}_2)/9.4\, \into 10^{20}$ mag, from \citet{bohlin78}  and the extinction law, \ak = 0.112 $\into$ \av, from \citet{rie85}, we find that \ak is related to \nht~ by  A$_\mathrm{K} = \mathrm{N({H}_2)}\, \into  1.2 \times 10^{-22}$ mag. Using this relation, we find that the opted \nht~threshold corresponds to \ak $\approx$ 0.2 mag, similar to the value opted for nearby GMCs for estimating cloud mass.

We use the following relation to convert the integrated column density ($\Sigma N(H_2)$) to  mass ($M_{\rm c}$):
\begin{equation}
M_{c} = \mu_{H_2} m_H A_{pixel} \Sigma N(H_2)
\end{equation}
where m$_H$ is the mass of hydrogen, $A_{pixel}$ is the area of pixel in cm$^2$, and $\mu_{H_2}$ is the mean molecular weight that is assumed to be 2.8 \citep{kauffmann2008}. Before integrating, we also subtracted a mean background \nht value of  $13 \times 10^{20}~ \cms$ from each pixel. This is done to correct for the contribution from  the diffuse material along  the line of sight. We estimated the mean background level from a relatively dust free region near the cloud.

We note, though the PPMAP provides a better resolution, but the output of the PPMAP technique can vary depending upon the variation in input parameters like opacity index and temperature bin resolution (e.g. PPMAP algorithm considers 12 temperature bins, equally spaced between 8 K and 50 K), which may give rise to uncertainty in column density and the estimated mass. \cite{marsh17} show that the global properties of a Hi-GAL field (i.e. a 2\degree.4 $\times$ 2\degree.4 tile of the Hi-GAL survey that hosts a molecular cloud M16) are not strongly affected due to variations in the input parameters. For example, the variation in mass is up to 20\%, and temperature is around $\pm$ 1 K,
for using $\beta$ in the range 2.0--1.5 and temperature bins from 12 to 8 K. To check the effects of the PPMAP assumptions on the global properties of  \cloud, we compared the maps of the PPMAP made by \cite{marsh17} with the maps of the \citet{schi2020}, made with the conventional method as described above. For Galactic plane clouds like \cloud, both the authors have used the images of the Hi-GAL survey, same opacity index ($\beta =2$) and gas-to-dust ratio ($R$ =100). We found that the properties of \cloud~largely remain the same, i.e. the difference in total mass is $\sim$ 15\% and in mean temperature is $\sim$ 2\%.  Although both methods give similar values of total mass, however, the true  uncertainty of mass can be high, 
as it depends on the number of properties such as dust opacity, gas-to-dust ratio, dust temperature, and distance.

In the present case, the dust temperature is unlikely the major cause of uncertainty for \cloud, but assuming an uncertainty of 30\% in dust opacity index and 23\% in the gas-to-dust ratio \citep[see][and discussion there in]{san2017} and using a distance  uncertainty of $\sim$ 9\%, we estimate the likely total uncertainty in our mass estimation to be around $\sim$ 45\%\footnote{It is worth noting that recent evidence shows that the gas-to-dust ratio varies with the Galactocentric radius \citep{Gian_2017}, although it is prone to large systematic error due to variations in the CO abundance and poorly constrained dust properties. If we opt for the gas-to-dust ratio value from the prediction of \cite{Gian_2017} for \cloud's galactic location, we find that the estimated mass
will be increased by a factor of 2.6. }.

Assuming circular geometry, we  calculated the effective radius as r$_{eff}$ = (Area / $\pi$)$^{0.5}$, the mean hydrogen volume density as n$_{\rm H_2}$ = 3M$_{c}$ / 4$\pi r_{eff}^3 \mu_{H_2} m_H$, and the mean surface density as $\Sigma_{gas}$ = $M_{c}$ / $\pi r_{eff}^2$ of the cloud. The total $M_{c}$, r$_{eff}$, the mean n$_{H_2}$, and the mean $\Sigma_{gas}$ for the cloud are found to be (1.1 $\pm$ 0.5) $\times$ 10$^5$\ \Ms, 26 pc, 22 $\pm$ 11 cm$^{-3}$, and 52 $\pm$ 25 \Ms~pc$^{-2}$, respectively. We find that these properties are consistent with 
those found in the Milky Way GMCs \citep[$\Sigma_{gas}$ = 50 \Ms~pc$^{-2}$, Mass $\geq$ 10$^{5-6}$ \Ms;][] {lada2020}. 

As discussed earlier, we also estimated the dense gas properties of \cloud~by integrating cloud area 
above \ak $\geq$ 0.8 mag. Doing so, we find the total $M_{c}$, the r$_{eff}$, the mean n$_{H_2}$, and the mean $\Sigma_{gas}$ to be 
(2.0 $\pm$ 0.9) $\times$ 10$^4$\ M$_\odot$, 6 pc, (3.21 $\pm$ 1.65) $\times$ 10$^2$ cm$^{-3}$, and (1.77 $\pm$ 0.85) $\times$ 10$^2$ M$_\odot$ pc$^{-2}$, respectively. These results are also summarised in Table \ref{tab:par}. We note that without background subtraction, the total mass and dense gas mass are 1.6 and 1.2 times higher than the mass measured with background subtraction. However, in the
present work, we have used the measurements estimated with the background subtraction. 

\begin{table*}

  \centering
  
   \begin{tabular}{c|c|c|c|c|c}
   \toprule
   \toprule
   \multicolumn{1}{c}{} & \multicolumn{2}{c}{\textbf{Dust continuum map}} & \multicolumn{2}{c}{\textbf{Extinction map}} & \multicolumn{1}{c}{} \\
    
   \cmidrule(rl){2-3} \cmidrule(rl){4-5}
    Parameter               & Whole Cloud  & Dense Gas & Whole Cloud & Dense Gas  & Unit    \\ \midrule
    Mass                   & (1.1 $\pm$ 0.5) $\times$ 10$^5$                        & (2.0 $\pm$ 0.9) $\times$ 10$^4$            & (9.1 $\pm$ 2.4) $\times$ 10$^4$        & (3.0 $\pm$ 0.8) $\times$ 10$^3$                     &  \Ms\\\\
    Effective Radius           & 26                        & 6            &24              & 2.4                     & pc     \\\\
    Average Volume density     & 22 $\pm$ 11                        & (3.21 $\pm$ 1.65) $\times$ 10$^2$        & 23 $\pm$ 8               & (7.52 $\pm$ 2.75) $\times$ 10$^2$            & cm$^{-3}$     \\\\
    Average Surface density    & 52 $\pm$ 25                        & (1.77 $\pm$ 0.85) $\times$ 10$^2$         & 50 $\pm$ 16        & (1.66 $\pm$ 0.52) $\times$ 10$^2$                        & \Ms~pc$^{-2}$    \\\\
    \bottomrule
        \end{tabular}\\
        
\caption{\cloud~properties from dust continuum and dust extinction maps.} 
\label{tab:par}

\end{table*}

\subsubsection{Properties from Near-infrared Extinction Map}
\label{ext}
One way to characterize the global properties of a star-forming cloud is to use its extinction map.
The advantage of using an extinction map in estimating column density is that it only depends on the extinction properties of the intervening dust, therefore providing an independent measure of cloud properties that can be compared with those obtained from dust continuum measurements. However, the limitation is that in the zone of high column densities where the optical depth in the infrared becomes too high to see background stars, it underestimates the column density values.

We generate a K-band extinction map  using the {\it UKDISS} point source catalogue, discussed in Section \ref{data} and implementing the PNICER algorithm discussed in \citet{mein2017}. The PNICER algorithm derives an intrinsic feature distribution along the extinction vector using a relatively extinction-free control field.  
It fits the control field data with Gaussian mixture models (GMMs) to generate the probability density functions (PDFs) that denote the intrinsic features, like intrinsic colours. The advantage of PNICER is that it uses all possible combinations of the near-infrared bands, such that the sources which do not have data in all wavelength bands will not affect the results. The PNICER creates PDFs for all combinations and automatically chooses the optimal extinction measurements for the target field \citep[for details, see][] {mein2017}. In the present case, for creating an extinction map of the \cloud~cloud, we choose a dust-free area close to the cloud area as our control field (i.e. the same area used for finding mean background column density).

Figure \ref{fig_dust}b shows the obtained K-band extinction map along with the CO contours. 
As can be seen, morphologically, the extinction map correlates well with the overall structure of the CO emission, as well as with the $Herschel$ column density map.
We find that within the cloud area defined by CO boundary, the dynamic range of our K-band extinction is in  the range 0.15 to 1.0 mag, with a median of 0.24 mag. The sensitivity limit of the extinction map is close to the sensitivity limit (i.e. \ak $\sim$ 0.2 mag) of the $Herschel$ column density map. 

Considering that the different approaches and tracers are used to make both the maps, the 
observed small difference at the cloud boundary is quite reasonable. However, we want to stress that unlike $Herschel$ map, our extinction map is 
insensitive to high column density zones of the cloud, which is the major source of uncertainty 
in estimating cloud properties, particularly the properties of the dense
gas content. In addition, the global properties of the cloud are also affected by systematic error in the adopted extinction law and distance. Here, we have taken the gas-to-dust ratio, $\frac{\mathrm{N(H}_2)}{\av}$ = $9.4\, \into 10^{20}$ cm$^{-2}$ mag$^{-1}$ based on a total-to-selective extinction, $R_V=3.1$ typical for the diffuse interstellar medium \citep{bohlin78}. However, the $R_V$ value can reach up to $\sim$ 5.5 \citep{chapman2009} in molecular clouds, for which the gas-to-dust ratio would change by $\sim$ 20\% \citep{cam1999}. For the \cloud~cloud, due to the combined uncertainties (i.e. due to extinction law and distance), the uncertainty in mass is around $\sim$ 27\%. This may be considered as lower-limit to the true uncertainty for clouds like $\cloud$ having high dense gas fraction
(discussed in Sect. \ref{dgf}). Nonetheless,
taking the estimated uncertainty as face value, we estimated the total $M_{c}$, the r$_{eff}$, the mean n$_{H_2}$, and the mean $\Sigma_{gas}$ for the \cloud~cloud as (9.1 $\pm$ 2.4) $\times$ 10$^4$\ M$_\odot$, 24 pc, 23 $\pm$ 8 cm$^{-3}$, and 50 $\pm$ 16 M$_\odot$ pc$^{-2}$, respectively. And for dense gas (\ak $\geq$ 0.8 mag), we find the total $M_{c}$, the r$_{eff}$, the mean n$_{H_2}$, and the mean $\Sigma_{gas}$ as (3.0 $\pm$ 0.8) $\times$ 10$^3$\ M$_\odot$, 2.4 pc, (7.52 $\pm$ 2.75) $\times$ 10$^2$ cm$^{-3}$, and (1.66 $\pm$ 0.52) $\times$ 10$^2$ M$_\odot$ pc$^{-2}$, respectively. All these measurements are also tabulated in  Table \ref{tab:par}.  As can be seen from the table,
the obtained dense gas properties are found to be lower than  the values obtained from the  column density map,
which we attribute to the fact that the inner area of the extinction map is not sensitive to the
high column density.

We note that, in general,  it has been found that the global properties of the cloud measured  from dust and extinction maps differ within a factor of 2--3 as both
the techniques involved different sets of assumptions \citep[e.g.][]{lom13,lom14,zari16}, all of which are difficult to evaluate independently. In this work, we do not intend to do a one-to-one comparison between the extinction map and dust column density map; rather, we are interested in the global properties of the cloud and its comparison  with the nearby clouds. 
Regardless of the different limitations of both
methods, in the present case, the global
properties of the whole cloud measured from both methods are in close agreement with each
other. This ensures the fact that the studied cloud is indeed a GMC of mass nearly 10$^5$ \Ms~enclosed
in a radius of $\sim$ 26 pc.  We also measured cloud mass from
CO analysis and found its total mass to be $\sim$ 0.8  $\times$ 10$^5$ \Ms. The detailed procedure of this measurement and a more thorough CO analysis of the cloud will be presented 
in future work. Here, our aim is to first report the global properties of the cloud using dust measurements.

\begin{figure*}
    \centering
    \includegraphics[width=12cm, height = 11cm]{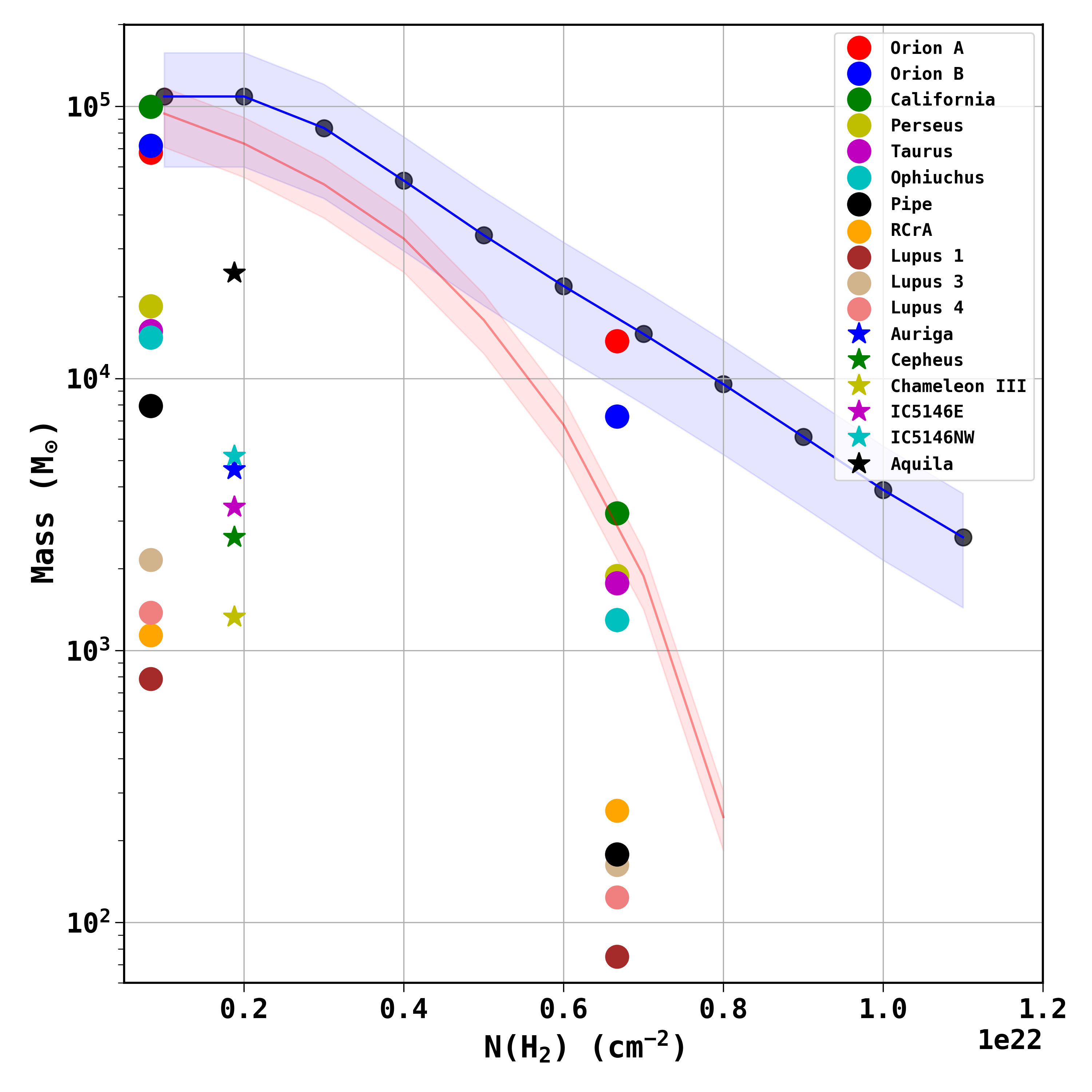}
    \caption{Enclosed mass of \cloud~at various column density thresholds. The blue and red lines show the cloud mass evaluated from the dust continuum and extinction map, respectively, at different column density thresholds. The shaded regions show the error in the estimated cloud mass. The coloured dots and stars show the mass of the nearby MCs taken from \citet{lada2010} and \citet{hei10}, respectively.  Only for putting Herschel and extinction based measurements at the same level, in this plot, we have extended the blue curve down to $\nht \sim 0.1 \times 10^{22}~ \cms$, however, since the Herschel column density map is not sensitive to column density less than $\sim 0.2 \times 10^{22}~ \cms$, the cloud mass remains flat. 
    }
    \label{fig_comp}
\end{figure*}

\begin{figure}
\centering
\includegraphics[width=8.5cm]{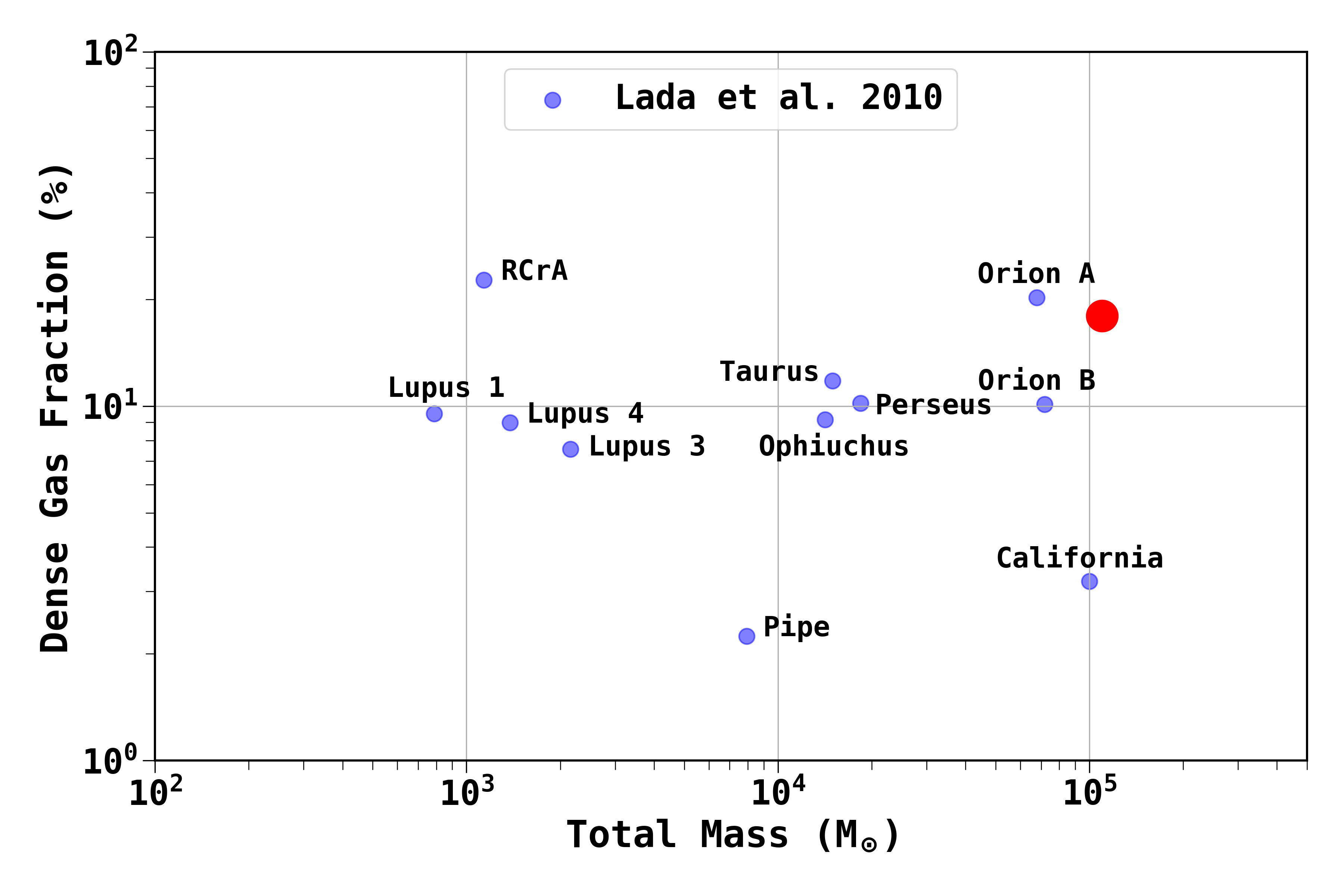}\\
\caption{Comparison of dense gas fraction of \cloud~with the nearby MCs given in \citet{lada2010}. The location of \cloud~is shown by a red circle.}
\label{fig_dgf}
\end{figure}

\subsubsection{Enclosed Mass and Dense Gas Fraction}
\label{dgf}

Figure \ref{fig_comp} shows the enclosed mass of the \cloud~cloud obtained from the $Herschel$ column density
map at different column density thresholds (shown by a solid blue line), and 
for the comparison purpose, we have also shown the cloud mass of the nearby GMCs as measured by \cite{lada2010} and \cite{hei10} at different column density  thresholds. As can be seen, the total mass of the 
\cloud~cloud is comparable  to the mass of the GMCs like Orion-A, Orion-B, and California and lies well above the mass of the other nearby molecular clouds. 
In Figure \ref{fig_comp}, we have also shown the \cloud~cloud's mass measured using the
extinction map at different thresholds (shown by a solid red line). As can be seen, from the extinction measurements also, the obtained total mass of \cloud~is comparable to the nearby GMCs. However, at the high-extinction threshold (e.g., \ak $\geq$ 0.8 mag), our measurements fall well
below the mass of Orion-A and Orion-B. This is because, unlike
nearby GMCs, our extinction map is not sensitive to the high column density zone of the \cloud~cloud. It is worth noting that the extinction maps used to measure the properties of the nearby GMCs are 
sensitive up to \ak $\sim$ 5 mag \citep[e.g. see Figure 1 of][] {lada2010}, while our map is sensitive up to \ak $\sim$ 1.0 mag. In general, the highest extinction that can be probed with the extinction  map is  sensitive to the distance of the cloud and the surface density of field stars in its direction.


As mentioned in Sect. \ref{gdp} for nearby clouds, a correlation exists
between the gas mass measured above the extinction threshold  of \ak$>$ 0.8 mag and the number of embedded YSOs identified in the infrared \citep{lada2010}.
Similar visual extinction thresholds in the range 7--8 mags are also
obtained by \citet{hei10,evans14} and \citet {andre14,andre2016} while analysing nearby GMCs using
extinction maps and dust column density maps, respectively. In particular,  \citet{lada12} found a strong linear scaling relation between star-formation-rate (SFR)
and dense gas fraction (i.e. $f_\mathrm{den}$= $\frac{ Mass (A_\mathrm{K}> 0.8\, mag)}{Mass (A_\mathrm{K} > 0.1 \, mag)}$). Therefore, the characterization of dense gas fraction is an important parameter for understanding the net outcome of star-formation processes, although it may not hold true for environments such as the Galactic Centre, where the critical density threshold for star formation is likely elevated due to the more extreme environmental conditions \citep{Hen_2022}.

 From dust analysis of \cloud, we find that the gas mass lies above \ak $\geq$ 0.8 mag is $\sim$ 2.0 $\times\, 10^4$ \Ms, while the total mass is $\sim$ 1.1 $\times\, 10^5$ \Ms, resulting
$f_\mathrm{den}$ as 18\%.  Figure \ref{fig_dgf} shows the comparison of dense gas fraction between \cloud~and the clouds studied by \citet{lada2010}. For nearby GMCs, these authors found the mean value of 
$f_\mathrm{den}$ as $0.10\pm 0.06$ with a maximum around 
$\approx 0.20$. 
As can be seen from Figure \ref{fig_dgf}, compared to nearby clouds, the $f_\mathrm{den}$ of \cloud~is on the higher side and comparable to that of the Orion-A, whose dense gas content is
$\sim 1.4 \times 10^4$ \Ms~\citep{lada2010}.
 We note, \citet{lada2010} measured, the total mass of Orion-A 
around $\sim 7 \times 10^4$ \Ms\ within an area of effective radius
$\sim$ 27 pc. In terms of total mass, effective area, and dense gas
mass, \cloud~resembles Orion-A. The above
analyses suggest that, like Orion-A, in \cloud, a significant fraction of mass is still in the 
form of dense gas.

We acknowledge that the caveat of this comparison is that it is drawn
by comparing measurements between the extinction map and the column density map. However,  as discussed above, the extinction maps of nearby clouds are sensitive up to \ak $\sim$ 5 mag, so we hypothesize that the high dense gas fraction that we observed in \cloud~may
hold true (or may not deviate significantly), if we had a more sensitive extinction map like nearby GMCs or if
the comparison is made with the $Herschel$ based measurements of nearby GMCs. To validate the later hypothesis, we measured the dense gas mass of 
Orion-A using the available $Herschel$ column density map of the Herschel Gould Belt Survey \citep{and10}, which is limited to the central area of the Orion-A cloud. 
Doing so, we find the total mass
of Orion-A to be $\sim 3 \times 10^4$ \Ms, while the total dense gas is $\sim 9 \times 10^3$ \Ms. 
The area covered by the $Herschel$ observations of Orion-A is less by a factor of two compared to the area covered by the extinction map used by \citep{lada2010}, thus, the estimated  total
mass from $Herschel$ is expected to be lower. However, we find
that  the dense gas mass obtained for Orion-A using both the aforementioned maps is largely the same. This is because the extinction map of Orion-A 
covers the entire dense gas area of the $Herschel$ map.

\begin{figure*}
\centering{
\includegraphics[width=8cm]{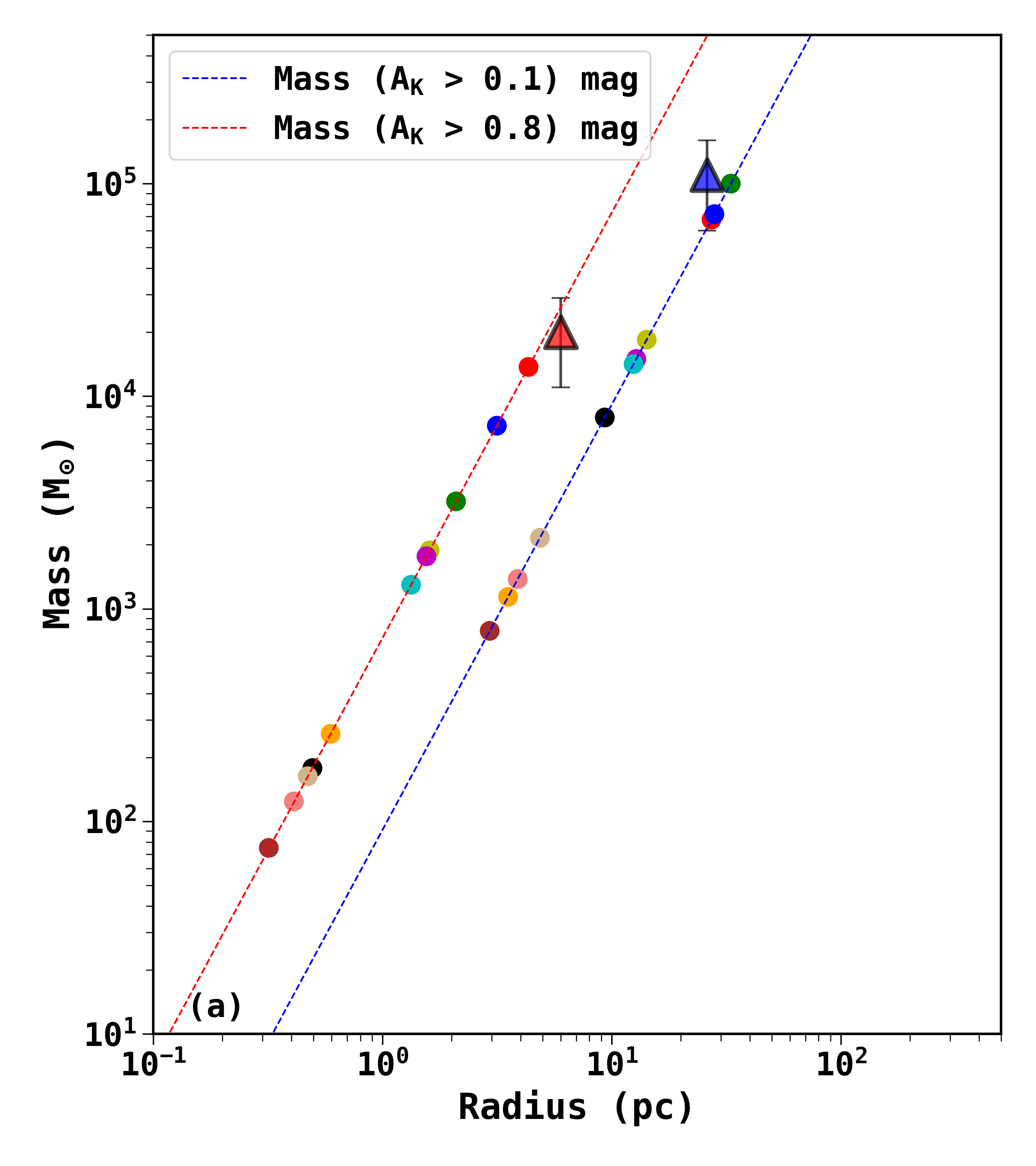}
\includegraphics[width=8cm]{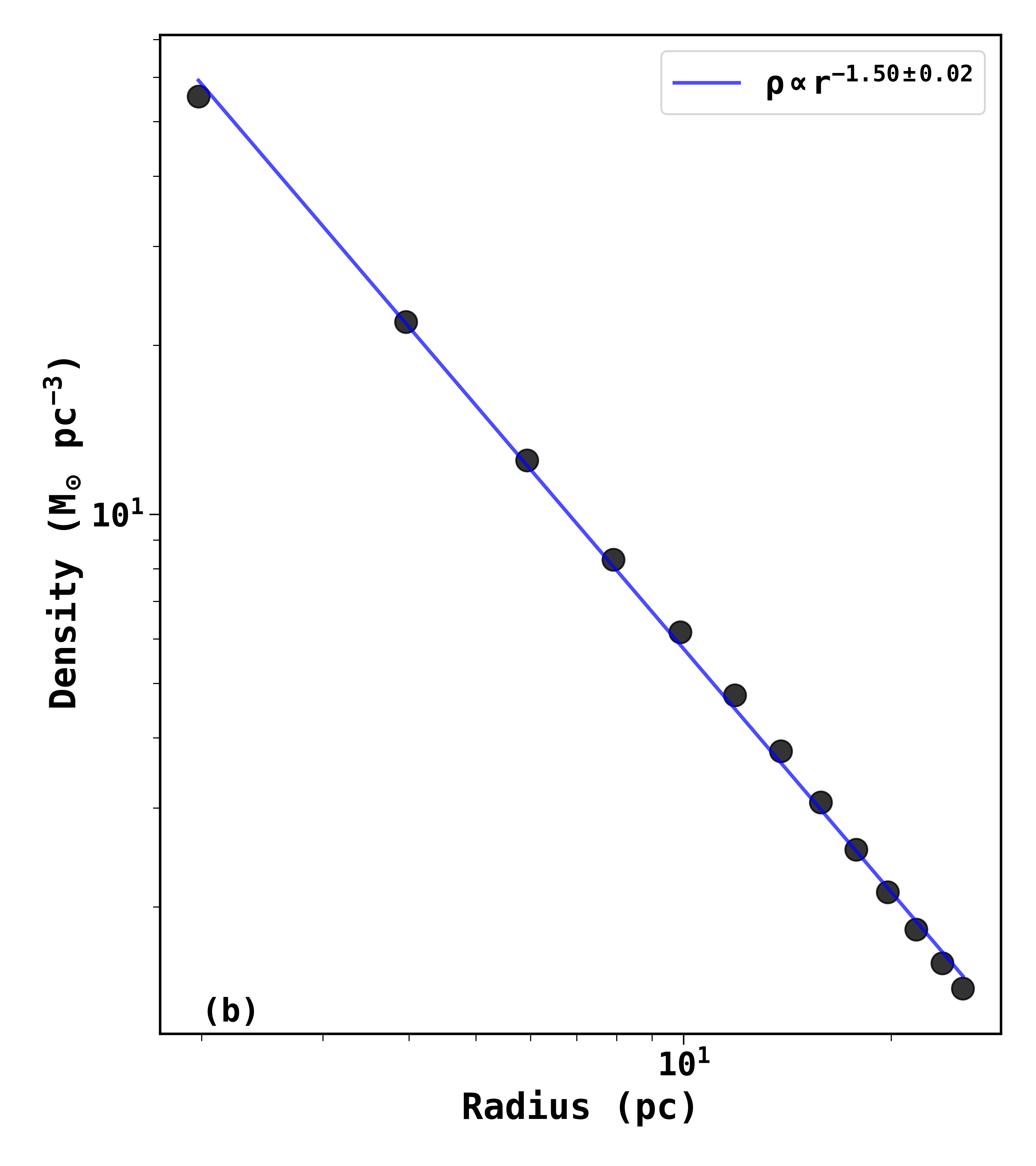}}
\caption{ (a) Cloud mass of the nearby molecular clouds estimated above extinction thresholds of \ak = 0.1 mag and \ak = 0.8 mag as a function of their radius. The color codes for the MCs are the same as shown in Figure \ref{fig_comp}, The blue and red dotted lines show the best-fitted mass-radius relation of the form, $M_{c} = \Sigma_{A_0}\pi R^2$, to the data for \ak = 0.1 mag and \ak = 0.8 mag, respectively. The location of \cloud~is represented by triangles with error bars.
(b) Density profile of \cloud~(shown by dots) along with the best-fitted power-law profile (shown by solid line) of 
index $\sim$ $-$1.50 $\pm$ 0.02.}
\label{fig_rdp}
\end{figure*}

\subsubsection{Structure and Density Profile}
\label{rdp}

The well-known scaling law, between the cloud size and mass,
$M_{c} = \Sigma_{A_0}\pi R^2$ was first documented by \cite{lar1981}.
\citet{hey09} using $^{13}$CO  observations (beam size $\sim$ 45\arcsec~and spectral resolution $\sim$ 0.2 \kms), estimated $\Sigma_{A_0}$ (mass surface density) value to be $\sim$ 42 $\pm$ 37 \Ms\ pc$^{-2}$ for larger
and distant GMCs.  
\citet{lom10} from their analysis of nearby clouds using extinction maps argued that $\Sigma_{A_0}$ depends on the
parameter $A_0$,  the extinction, defining the outer boundary 
of the cloud and 
found that $\Sigma_{A_0}$ value increases as the extinction threshold increases.  They also suggested that all clouds follow a Larson-type relationship, therefore, very similar projected mass densities at each extinction threshold. However, they find that  the
mass-radius relation for single clouds does not hold in their sample, indicating that individual clouds are not objects that can be described by constant column density. 

In Figure \ref{fig_rdp}a, we compare the mass-size relation of \cloud~measured from the dust column density map with the data of the nearby clouds studied by \cite{lada2010}. The dotted lines show the  least square fit of the form $M_{c} = \Sigma_{A_0}\pi R^\gamma$ with $\gamma \sim$ 2 to the measurements of the nearby clouds \citep{lada2010, krum-dek_2012} at \ak $\geq$ 0.1 mag  and \ak $\geq$ 0.8 mag. By doing so,
we estimated the $\Sigma_{A_0}$ value to be 29.1 $\pm$ 0.1 \Ms\ pc$^{-2}$ and 232.1 $\pm$ 0.1 \Ms\ pc$ ^{-2}$ for mass measured above \ak $\geq$ 0.1 mag  and \ak $\geq$ 0.8 mag, respectively. 
Despite different methods being used for mass measurements, as can be seen from Figure \ref{fig_rdp}a, the \cloud~cloud (shown by large triangles) closely follows the mass-size relation of the nearby clouds for different thresholds. For \cloud,  the slightly high $\Sigma_{A_0}$ corresponding to scaling-law \ak $\geq$ 0.1 mag could be due to the fact that its mass has been measured at a higher extinction threshold. i.e. at \ak $\geq$ 0.2 mag, as our $\it{Herschel}$ column density map is not sensitive below \ak= 0.2 mag.
This figure also shows that the dense gas and total mass of \cloud~are higher than the ones for the nearby GMCs.

The density profile of a single cloud is also important for theoretical considerations of star formation. For example, it has been suggested that a density profile of the form, $\rho$(r) $\propto$ r$^{-1.5}$, is indicative of a self-gravitating spherical cloud  supported by turbulence \citep [e.g.][]{mur15}, while a profile of the form, $\rho$(r) $\propto$ r$^{-2}$,  is indicative of a gravity dominated system \citep[e.g.][]{don18,li18,chen21}.  Figure \ref{fig_rdp}b shows the density profile of the \cloud~cloud
along with the best-fitted power-law  profile (blue solid line). We find that $\rho$(r) $\propto$ r$^{-1.5}$ best fits the overall large-scale structure of the cloud. We note that this is the overall density profile of the cloud, but individual dense pockets of gas or clumpy structures can have a steeper density profile. As \cloud~is located at 3.4 kpc, unveiling the profile of such structures would require high-resolution observations. However, it is worth mentioning that steeper density profiles with an average power-law 
index between $-$1.8 to $-$2 have been observed in massive star-forming clumps \citep[e.g.][]{gary07}.

\label{pro_con}

\begin{figure*}
\centering{
\includegraphics[width=17.9cm]{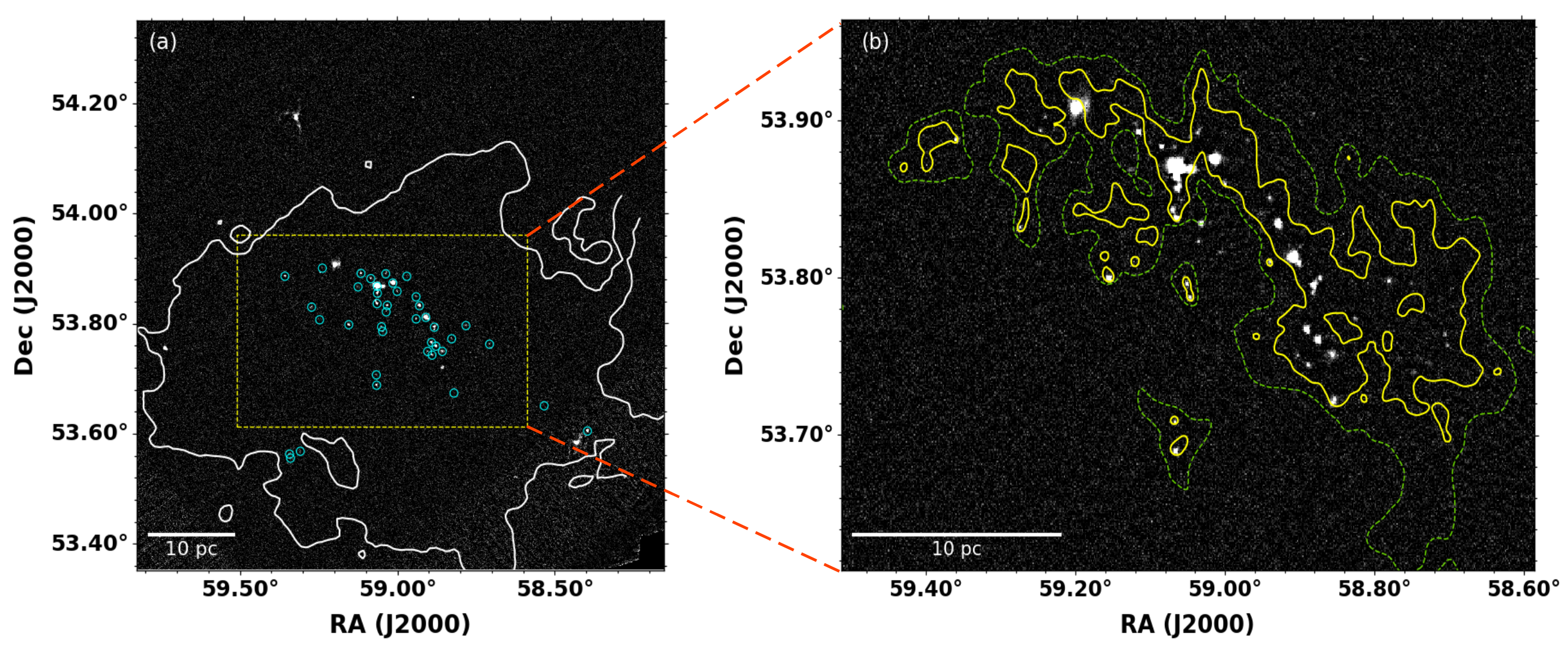}}
\caption{(a) Unsharp-masked 70 $\mu$m $Herschel$ image of \cloud~over which 70 $\mu$m point sources are highlighted in cyan circles. The white contour shows the outermost contours of  CO integrated emission. (b) A zoomed-in view of the central area of the cloud. The green dotted and yellow solid contour corresponds to the column density value of 5.0 $\times$ 10$^{21}$ cm$^{-2}$ (\av $\sim$ 5 mag) and 6.7 $\times$ 10$^{21}$ cm$^{-2}$ (\av $\sim$ 7 mag), respectively, enclosing the distribution of most of the  70 $\mu$m point sources.}
\label{fig_pac}
\end{figure*}

\subsubsection{Boundness Status of G148.24+00.41}
\label{bs}
A large reservoir of bound gas is key for making massive clusters because theories and simulations often invoke gas inflow from large scale \citep[e.g.][]{gom14,pad2020}.  However, it is suggested that on relatively short time-scales (typically a few Myr) since its formation, GMCs can be unbound as colliding flows and stellar feedback regulate the internal velocity dispersion of the gas and so prevent global gravitational forces from becoming dominant \citep{dobbs11}. So relatively older clouds can be unbound, although it is also suggested that large-scale unbound clouds can also have bound substructures where star cluster formation can take place \citep{clark-bon2004, clark2005}.

Whether a cloud is bound or not is usually addressed
by calculating the virial parameter, $\alpha=\frac{M_{vir}}{M_c}$, which compares the virial mass to the actual mass of the cloud \citep{ber-mck1992}.
A cloud is bound if  $\alpha_{vir}$ $<$ 2  and unbound if  $\alpha_{vir}$ $>$ 2 \citep{mao2020}. It is in gravitational equilibrium if its actual mass and virial mass are equal.

We estimated the virial mass for \cloud, assuming it is a spherical cloud with a density profile, $\rho \propto r^{-\beta}$, and using the
equation from \citet{MacLaren1988} in the following rewritten form:
\begin{eqnarray}
M_{\rm vir} (M_\odot) = 126 \left(\frac{5-2\beta}{3-\beta}\right) \left(\frac{R}{{\rm pc}}\right) \left(\frac{\Delta \rm{V}}{{\rm km~s}^{-1}}\right)^2, 
\end{eqnarray}
where $R$ is the radius of the cloud and $\Delta \rm{V}$ is the line-width
of the  gas. Here, we adopt $\beta=1.50 \pm 0.02$
and $\Delta \rm{V}$ = 3.55 $\pm$ 0.05 \kms (see Section \ref{dis}) and assume
that $\Delta \rm{V}$ describes the average line width of 
the whole cloud, including the central region. The derived virial mass turns out to be $\sim (5.50 \pm 0.16) \times 10^4$ \Ms, while the estimated dust mass
is $\sim (1.1 \pm 0.5) \times 10^5$ \Ms~(see Section \ref{dust}), resulting 
$\alpha \simeq 0.5 \pm 0.2$, and hence the whole cloud is likely bound. This remains true even if we use the lower mass obtained from extinction analysis.

\begin{figure}
\centering{
\includegraphics[width=8.5cm]{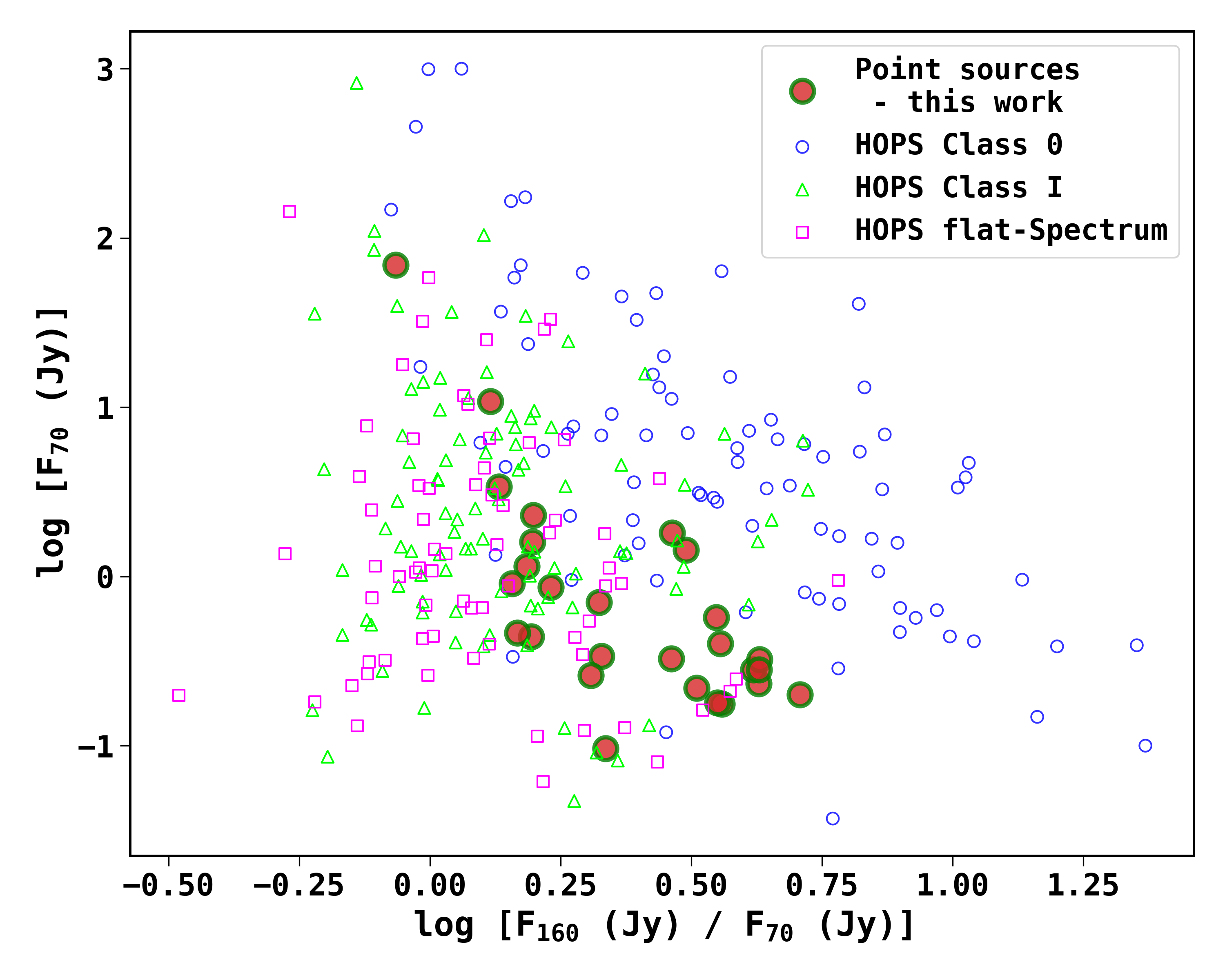}}
F\caption{Plot of 70 $\mu$m flux density against 160 $\mu$m to 70 $\mu$m flux density ratio for protostars (shown by red solid circles) of the \cloud~cloud. In the plot, the open circles, triangles, and squares are the class 0, class I, and flat-spectrum sources, respectively,  from the $Herschel$ Orion Protostar Survey.}
\label{fig_hops}
\end{figure}

\subsection{Protostellar Content and Inference from their Distribution}

\subsubsection{$Herschel$ Point Sources and their Evolutionary Status}
\label{hps}
The $Herschel$ satellite offers a unique opportunity to study the earliest phases of stellar sources. In particular, $Herschel$ 70 $\mu$m band is very important for identifying deeply embedded class 0/I objects because it has been found that 70 $\mu$m is less sensitive to circumstellar extinction and geometry of the disc that significantly affects the 3.6--24 $\mu$m band. 70 $\mu$m is also less affected by external heating that becomes effective above 100 $\mu$m \citep{dun2006}.

Figure \ref{fig_pac}a shows the $Herschel$ 70 $\mu$m image of the \cloud~cloud. As can be seen, the image displays a significant number of sources distributed roughly in a linear sequence from north-east to south-west, and most of the sources seem to be embedded in high column density material of $\nht > 5.0 \times 10^{21} \cms$. Since such high column density regions of a molecular cloud are the sites of recent star formation \citep[e.g.][]{and10}, these sources are possibly young protostars of the \cloud~cloud at their early evolutionary stages.

To understand the nature of the sources, we downloaded the $Herschel$ 70 $\mu$m point source catalogue \citep{marton2017, hers2020} from Vizier \citep{viz2000}. 
In total, we retrieve 48 point sources within the cloud radius having SNR $>$ 3.0.
As noted by \citet{hers2020}, the  detection limit of the $Herschel$ point source catalogue is a complex function of the source flux, photometric band, and the background complexity. We thus look to the reliability of the downloaded
point sources by visually inspecting their 
positions and intensities on the 70 $\mu$m image.
We find that some sources are too faint to be considered as a point source, and also find that
a few likely bright sources (which appear to be extended on the image) are also missing in the catalogue. The former could be the artefact due to the non-uniform background level usually found in $Herschel$ images, while the latter 
could be due to the fact that these sources failed to pass the point source quality flags such as confusion and blending flags, implemented
in  \citet{hers2020}
to be called a point source. To check the reliability of the faint sources, we create different unsharp-masked images by subtracting median-filtered images
of different windows from the original one \citep[e.g.][]{deh15}. Unsharp-masked images are useful to detect faint sources or faint structures that are hidden inside bright backgrounds. We again over-plotted the point sources and found that a few sources  are likely false detections, thus, did not use them in this work. After removing likely spurious sources, the total number of point sources is 40, with the faintest being a source of $\sim$ 96 mJy. In Figure \ref{fig_pac}a, these point sources are marked in cyan circles. 

In order to assess the evolutionary status of the point sources, we compared their location
in the 70 $\mu$m flux density versus 160 $\mu$m to 70 $\mu$m flux density ratio diagram with that of the well-known protostars of the Orion complex, shown in Figure \ref{fig_hops}. We took the Orion protostar sample from the $Herschel$ Orion Protostar Survey \citep[HOPS;][]{fur2016}. It consists of 330 sources that have 70 $\mu$m detection, 319 of which
have been classified as  class 0, class I, or flat-spectrum protostars based on
their mid-IR spectral indices and bolometric temperatures, while 11 sources have been classified as class II objects. 
As can be seen from Figure \ref{fig_hops}, most of the point sources  (red
dots) have 160 $\mu$m to 70 $\mu$m flux density ratio $\geq$ 1.0 like the HOPS protostars. 
The only source that shows a relatively smaller ratio with respect to the rest of the 
sources is the most luminous 70 $\mu$m source. This source is  the most massive 
YSO in our sample (more discussion in Section \ref{lf} and \ref{ms}).
To access the degree of contamination that might be present
in our sample, in the form of extragalactic sources or other dusty objects
along the line of sight, we did a similar analysis for the point sources
present in the control field region. We find that none of the control field sources are located in the zones of the 
HOPS protostars, implying that contamination is negligible and the majority of the identified 70 $\mu$m point sources
in \cloud~are likely true protostars. The identification of these protostars suggests that, the central area of the
cloud is actively forming protostars compared to the rest of the cloud.

\begin{figure}
\centering{
\includegraphics[width=8.5cm]{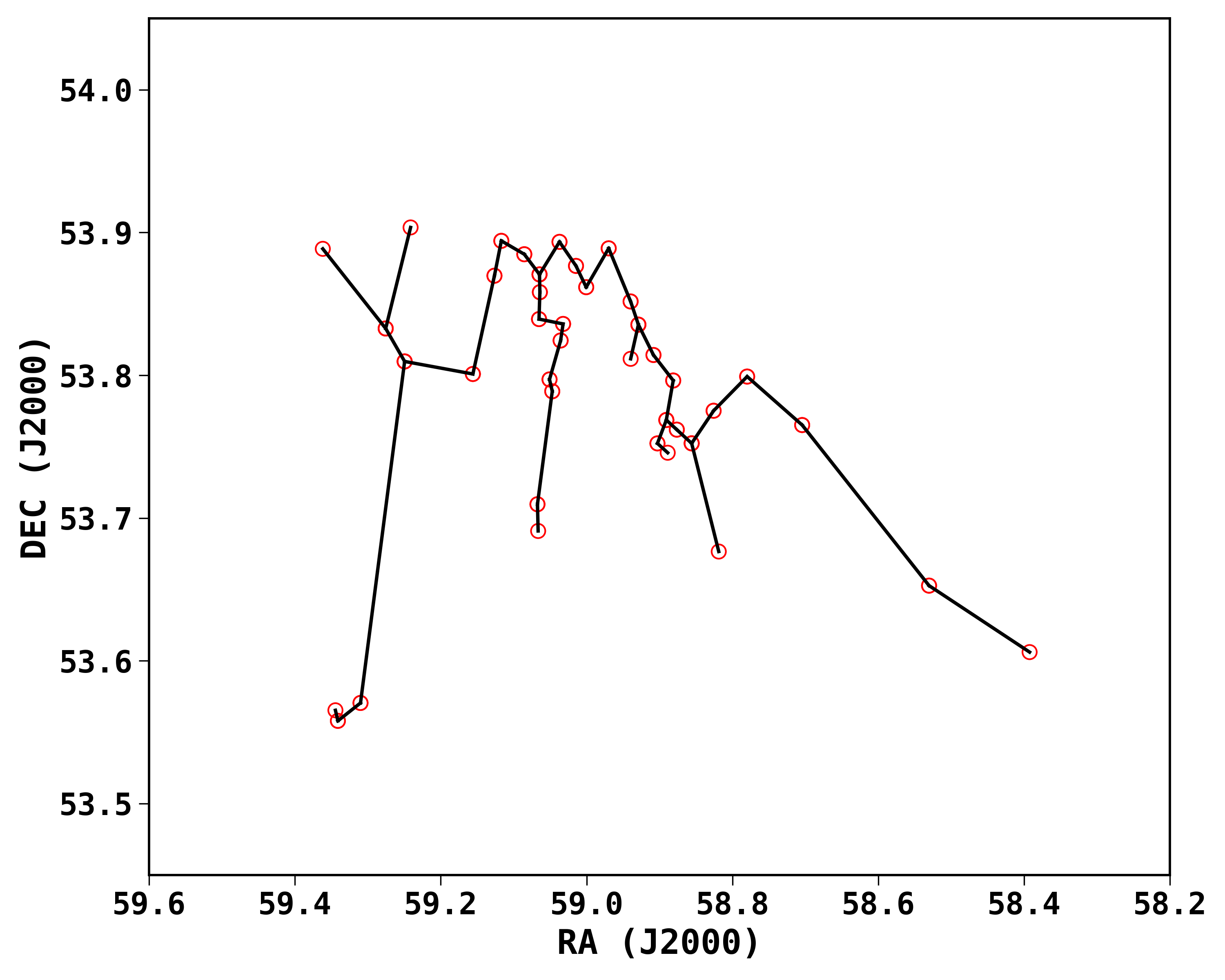}}
\caption{Minimum spanning tree distribution of the protostars in our sample. The red circles indicate the positions of the protostars, while the lines denote the spanning edges.}
\label{fig_mst}
\end{figure}

\subsubsection{Fractal Nature of the Cloud}
\label{fs}
For clouds and clumps at their early stage of evolution, the distribution of cores or young protostars carries the imprint of the original gas distribution.

We examine the structure of the \cloud~cloud using 
$Herschel$ identified protostars and implementing 
the statistical Q-Parameter method \citep{cart-whit2004}, which is based on the minimum spanning tree (MST) technique.  A MST is defined as the unique network of straight lines that 
can connect a set of points without closed loops, such that the sum of all the lengths of these lines (or edges) is the lowest. The Q-Parameter method has been extensively studied in
the literature for understanding the clustering (large-scale radial density gradient or small-scale fractal) structure of the star-forming regions and molecular clouds 
\citep[e.g.][]{sch-kle2006, park2014, san2019, dib-hen2019, sad2020}. 
Q is expressed by the following equation:
\begin{equation}
    Q = \frac{ \bar{l}{_{edge}}}{\bar{s}},
\end{equation}    
where the parameter $\bar{l}{_{edge}}$ is the normalized mean edge length of MST, defined by:
\begin{equation}
\centering {
\bar{l}{_{edge}} =\frac{(N-1)}{(A N)^{1/2}} \sum_{i=1}^{N-1}{m_i}
}
\end{equation}
where $N$ is the total number of sources,
$m_i$ is the length of edge $i$, and $A$ is the area of the smallest circle that contains all
the sources. The value of $\overline{s}$ represents the correlation length, i.e. mean projected separation of the sources normalized by the cluster radius and is given by
\begin{equation}
\centering
\overline{s} = \frac{2}{N(N-1)R} \sum_{i=1}^{N-1}{} 
\sum_{j=1+i}^{N} \vert \overrightarrow{r}_i - \overrightarrow{r}_j \vert
\end{equation}
where $r_i$ is the vector position of the point $i$ and $R$
is the radius corresponding to area A. The $\bar{s}$ decreases more quickly than $\bar{l}{_{edge}}$ as the degree of central concentration increases, while $\bar{l}{_{edge}}$ decreases more quickly than $\bar{s}$ as the degree of subclustering increases \citep{cart-whit2004}. Therefore, the Q parameter not only quantifies but also differentiates between the radial density gradient and fractal subclustering structure. 

Figure \ref{fig_mst} shows the MST graph of the protostars in our sample. In the present case, we define the radius as the projected distance from the mean position of all cluster members to the farthest protostar following \cite{sch-kle2006}. Doing so, we calculated $\bar{l}{_{edge}}$ and  $\bar{s}$ as 0.27 and 0.41, respectively, which leads to a  Q value of $\sim 0.66$. Including the protostars identified by 
the Star Formation in the Outer Galaxy (SFOG) survey \citep{wins2020}, that has used
{\it Spitzer-IRAC}, {\it WISE}, and {\it 2MASS} data in the wavelength range 1-22 
 $\mu$m, though we improved the statistics of protostars sample to 70, but find that Q-value largely remains the same, i.e. Q $\sim$ 0.62, a change of only 6\%. 

We note that the normalization to cluster radius makes the Q parameter scale-free, but 
a small dependence of the number of stars on the Q parameter is found in simulations \cite[e.g.][]{Park2018}. This is also seen in our analysis, as we found a change in Q value only by 6\%. Apart from the number of stars, the presence of outliers can also significantly affect the Q parameter \citep{Park2022}. To check the significance of outliers on our Q value, we removed possible outliers from our sample, which are far away from the main star-forming region (i.e. sources located outside the rectangular box shown in Figure \ref{fig_pac}a; these sources are also located away from the \av  $\sim$ 5 mag contour, shown in the right panel of Figure \ref{fig_pac}b) and did the MST analysis. We found that the Q value changes only by 3\%, resulting a total uncertainty of $\sim$ 7\% (i.e. Q $= 0.660 \pm 0.046$) due to the above factors. 

We also looked at how the completeness of the 70 $\mu$m catalogue could affect our Q-value. Since most of the protostars are distributed in the central
area of the cloud with \av$>$ 5 mag, we estimated the completeness limit
of the 70 $\mu$m catalogue in the central area, i.e. the area roughly enclosing the boundary of the \av$>$ 5 mag. We estimated
completeness by injecting
artificial stars on the 70 $\mu$m image
and performing detection and photometry in the same way as done in the original catalogue \citep[for details, see][]{marton2017}. Doing so, we find that our 70 $\mu$m point source sample in the central
area is $\sim$80\% complete  at the flux level of $\sim$ 200 mJy. Recalculating the Q value above the 80\% completeness limit, we find the Q value to be around 0.71.

The value of Q $>$ 0.8 is interpreted as a smooth and centrally concentrated distribution with
volume density distribution $\rho(r) \propto r^{\alpha}$, while Q $<$ 0.8 is interpreted as clusters with fractal substructure, and Q $\simeq$ 0.8 implies uniform number density and no subclustering \citep[see][for discussion]{cart-whit2004}. \citet[][]{cart-whit2004} drew these inferences  by studying
 the structure of the artificial star clusters, created with a smooth large-scale radial density profile ($\rho(r) \propto r^{\alpha}$) and with substructures having fractal dimension D, and correlating them with the Q-value. The
 fractal substructures of various fractal dimensions were generated following the  box fractal method of \citet{good04}. We direct the reader to \citet{good04} for the details of the box fractal method. \citet {cart-whit2004} found that Q is correlated with the radial density exponent $\alpha$ for
 Q $>$ 0.8, and for fractal clusters, the Q is related to fractal dimension D such that the  Q parameter changes from 0.80 to 0.45 as the D changes from 3.0 (no subclustering) to 1.5 (strong subclustering).  From the simulation results
 of \citet[][shown in their Fig. 5]{cart-whit2004}, we infer that our  estimated Q value (i.e. $Q = 0.660 \pm 0.046$), corresponds to a notional fractal dimension, D $\sim$ 2.2, which represents a moderately fractal distribution. 




Since most of the protostars are distributed in the central area of the
cloud, shown in Figure \ref{fig_pac}. From this analysis, we infer that the cloud in its central area is moderately  
fractal.

\subsubsection{Luminosity of Protostars and their Correlation with the Gas Surface Density}
\label{lf}

\begin{figure}
\centering{
\includegraphics[width=8.2cm]{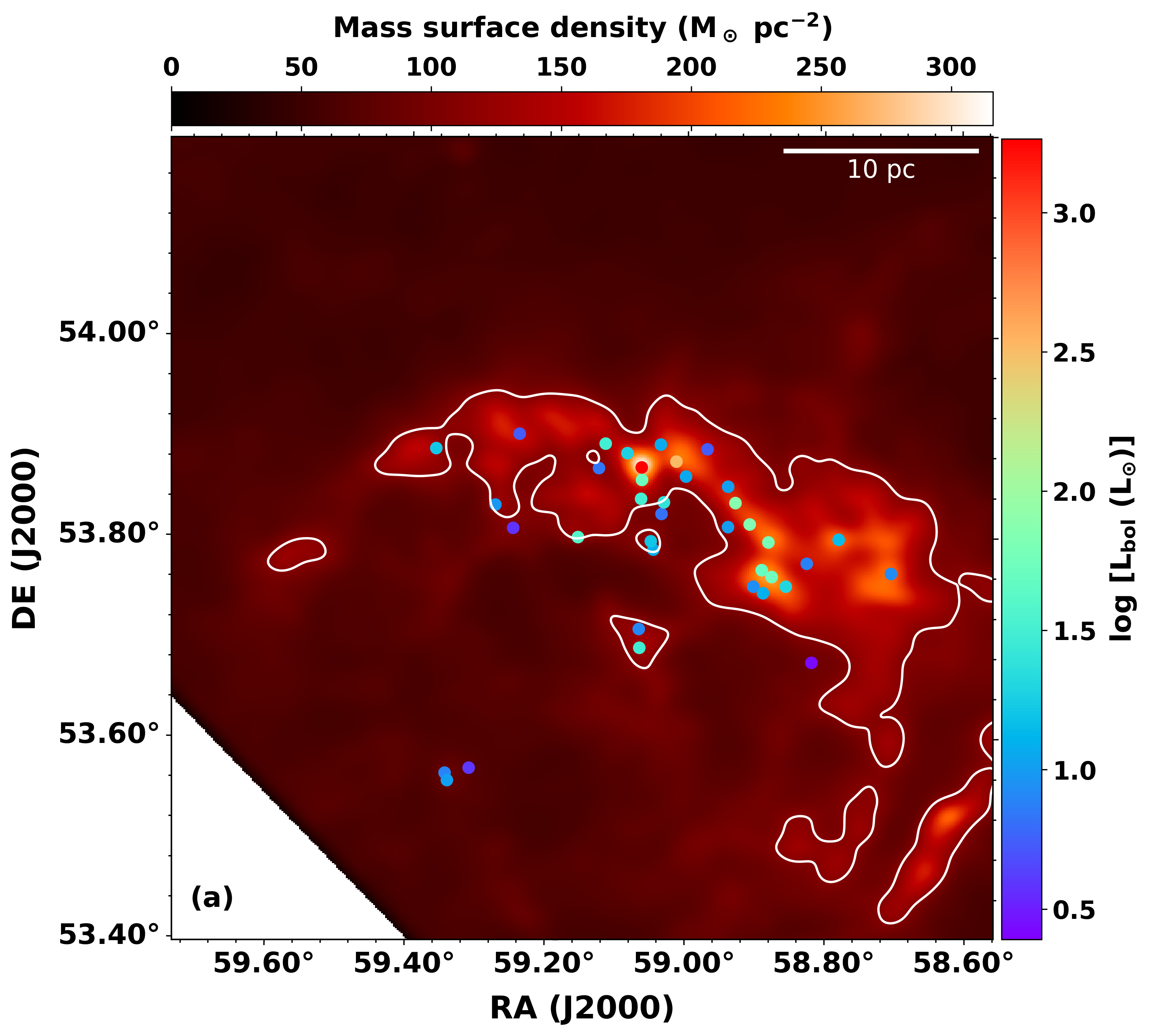}
\includegraphics[width=8.4cm, height = 6.4cm]{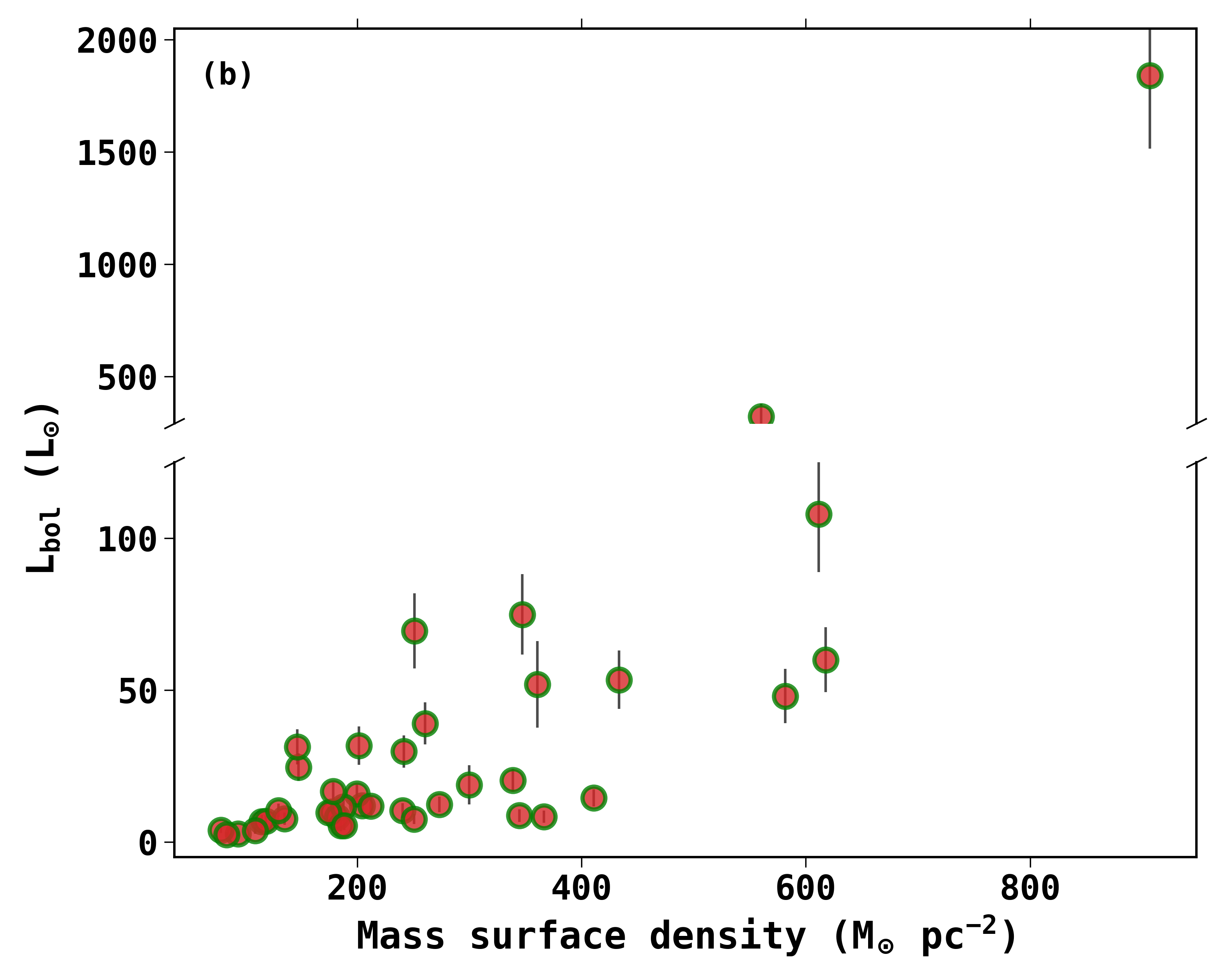}
\includegraphics[width=8.4cm, height = 6.4cm]{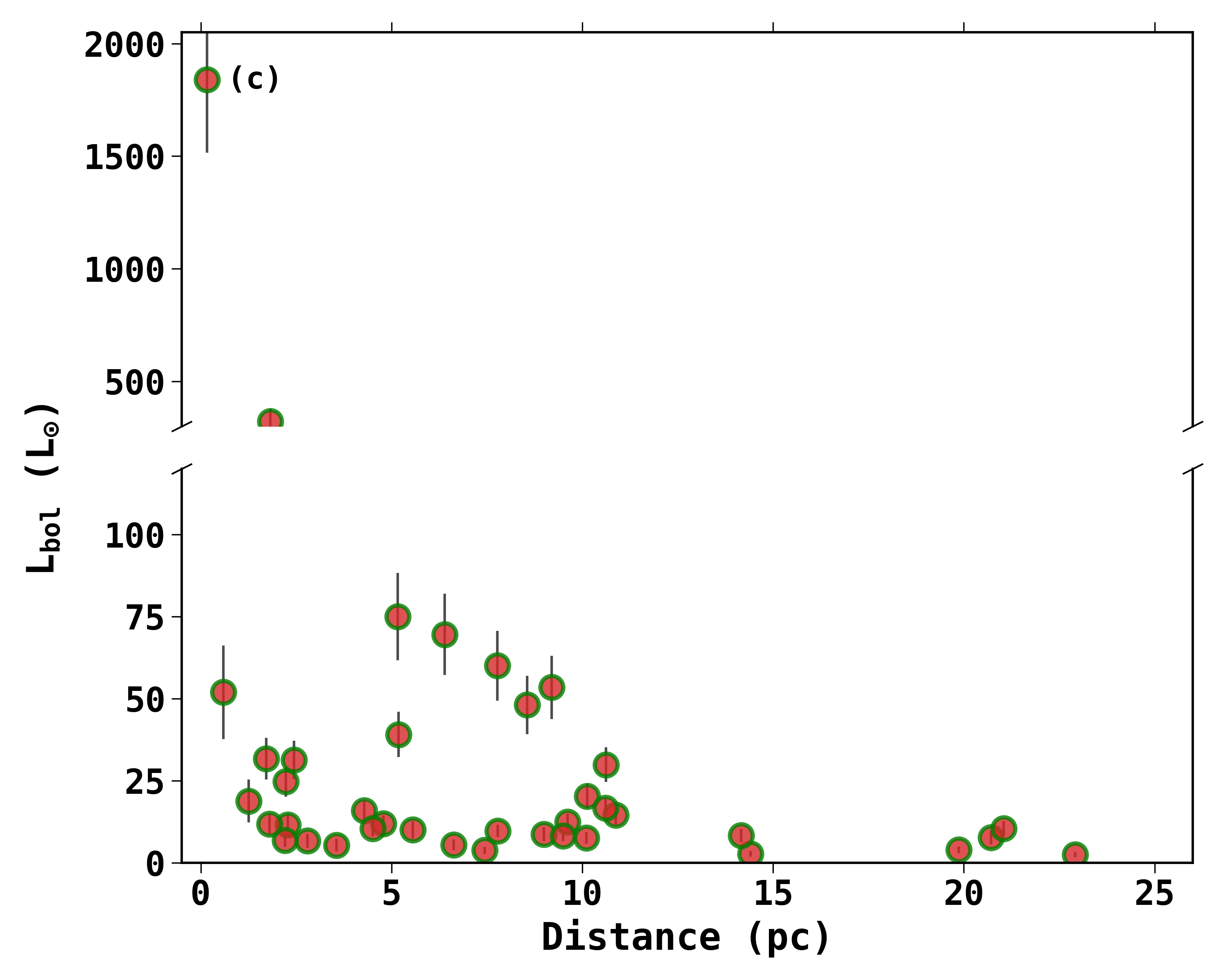}
}
\caption{(a) Spatial distribution of protostars on the smoothed mass surface density map (beam $\sim$ 30\arcsec). The overplotted white contour corresponds to the mass surface density of 110 \Ms~pc$^{-2}$ that encloses most of the sources. The right colorbar shows the bolometric luminosity of the protostars on a log-scale. The highest luminosity source (\lbol $\approx$ 1900 \lsun) is shown by a red dot. (b) Plot showing the luminosity of the protostars vs. their corresponding mass surface density.
(c) Plot showing the radial distribution of luminosity of the protostars from the likely centre of cloud's potential.}
\label{fig_lbol}
\end{figure}

\cite{dun2006}, using radiative transfer models, demonstrated that 70 $\mu$m is a crucial wavelength for determining bolometric luminosity (\lbol) of embedded protostars, as radiative transfer models are strongly constrained by this wavelength, and it is largely unaffected by the details of the source geometry and external heating. Furthermore, \cite{dun2008} and \cite{rag2012} find that the 70 $\mu$m flux correlates well with the bolometric luminosity of the low and high-mass protostars, respectively \citep[see also discussion in ][]{eli17}. 

We find that a significant number of the 70 $\mu$m point sources are blended at longer $Herschel$ wave bands. Thus, in this work, we used the empirical relation between 70 $\mu$m flux and \lbol given by \cite{dun2008} for estimating \lbol of the protostars:

\begin{equation}
    L_{bol} = 3.3 \times 10^8 F_{70}^{0.94} \left(\frac{d}{140 pc}\right)^2 L_{\odot},
\end{equation}

where $F_{70}$ is in erg cm$^{-2}$ s$^{-1}$, though  this way of estimating luminosity is likely accurate within a factor of 2–3 \citep[e.g.][]{com2012, samal2018}.
We estimated the luminosity of the sources keeping this limitation in mind and found that they lies in the range $\sim$ 3--1850 \lsun, with a  median value around  12 \lsun. 

Figure \ref{fig_lbol}a shows the luminosity distribution of the sources on the mass surface density map, made from the column density map. The uncertainty in the luminosity is due to the uncertainty in the distance of the cloud and the flux of the point sources. As can be seen that the most luminous source (the reddest solid dot)  is located in the zone of the highest 
surface density, and most of the sources are confined to surface density $>$ 110 \Ms~pc$^{-2}$. 
Figure \ref{fig_lbol}b shows the bolometric luminosity versus peak mass surface 
density corresponding to the source location. From the figure, one can see 
that sources are distributed in the surface density range 80--900 \Ms~pc$^{-2}$ and shows  a positive correlation with the mass-surface density, implying that higher luminous sources are found in the higher surface density zones. 
Figure \ref{fig_lbol}c shows the luminosity distribution of the sources from the location of the cloud's likely centre of potential. As discussed in Section \ref{dust}, the inner region of the cloud is elongated and filamentary, thus, finding its centre of gravitational potential is not easy, so we define the cloud's 
gravitational centre  as the location of the highest surface density area on the smoothed surface density map (shown in Figure \ref{fig_lbol}a).  We made a smoothed map to understand the structure that dominates the large-scale distribution as a function of scale. The highest surface density area on the smoothed map also  corresponds to the location of a hub, seen in the $Herschel$ SPIRE images (discussed in Section \ref{hub}). From Figure \ref{fig_lbol}c, a declining trend of luminosity distribution with the distance from the adopted centre can be seen, although many low-luminosity sources are also located close to the centre along with the most luminous source.

Altogether, the above analyses suggest that although protostars are distributed in a range of surface densities, the  luminous sources
are located in the highest surface density zones and also close to the cloud's centre of potential.

\begin{figure}
\centering{
\includegraphics[width=8.5cm]{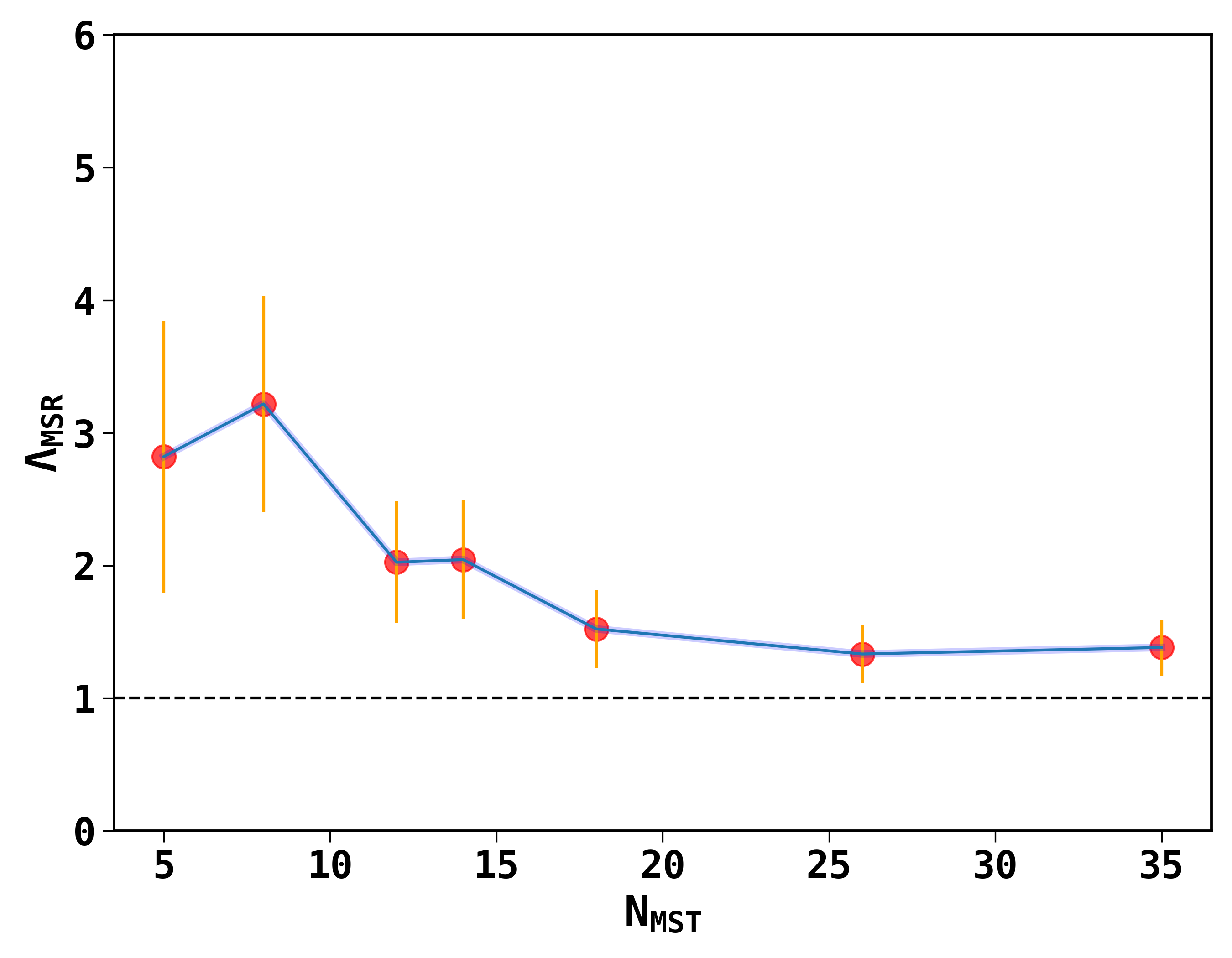}}
\caption{Plot showing the evolution of $\mathrm{\Lambda_{MSR}}$ for \cloud with different number of most massive sources, N$\mathrm{_{MST}}$. 
 The dashed line at $\mathrm{\Lambda_{MSR}}$ $\sim$ 1 shows the boundary at which the distribution of massive stars is comparable to that of the random
stars.}
\label{fig_msr}
\end{figure}

\subsubsection{Mass Segregation}
\label{ms}

A higher concentration of massive objects near the cloud or cluster centre compared to that of their low-mass siblings is known as mass-segregation. However, it remains unclear whether mass segregation is primordial or dynamical, particularly in young 
star-forming regions or cluster-forming clumps. Mass segregation is an important constraint on theories of massive stars and associated cluster formation. For example, it is suggested that cores in the dense central regions of cluster forming clumps can accrete more material than those in the outskirts; therefore, primordial mass segregation would be a natural outcome of massive star formation \citep{bon-bate2006, gir2012}. However, it is also suggested
that mass segregation can also occur dynamically. In this scenario, massive stars form elsewhere but 
sink to the centre of the system potential through dynamical interaction with the other members
\citep[e.g.][]{alli10, park2014,park16,dom2017}.
 
One way to test the above theories is to look for the distribution of young protostars and cores in  young molecular clouds. Because, the velocity dispersion of cores in young star-forming regions is found to  be  $\sim$ 0.5 \kms, while  the velocity dispersion of the class II sources in the same regions is found to be  higher, at around 1 \kms \citep[e.g. NGC 1333;][]{fost2015}. Thus, the class II stars of a star-forming region can travel pc-scale distance in a Myr timescale from their birth locations, while protostars being young (age $\sim 10^5$ yr) and often attached to the host core, nearly represent their birth locations. 

To quantify the degree of mass-segregation ($\Lambda_{MSR}$) in star-forming regions, \cite{ali2009} described a statistical way that uses MST distribution of stars. This method works by comparing the average MST length of the most massive stars of a cluster with the average MST length of a set of the same number of randomly chosen stars and is written as:
\begin{equation}
\Lambda_{MSR} (N) = \frac{<l_{random}>}{l_{massive}} \pm \frac{\sigma_{random}}{l_{massive}},
\end{equation}
For good statistical results, one needs to take a significant number of random samples \citep{mas-clark2011}. Here, $<l_{random}>$ is the sample mean of average MST lengths of $N$ randomly selected stars, $l_{massive}$ is the average MST length of $N$ most massive stars, and $\sigma_{random}$ is the standard deviation of the length of these $N$ random stars.  
The $\Lambda_{MSR}$ greater than 1 means that the $N$ most massive stars are more concentrated compared to a random sample, and therefore, the cluster shows a signature of mass segregation, while  $\Lambda_{MSR} \sim 1$ implies that the distribution of massive stars is comparable to that of the random stars.

In the present work, we have not estimated the mass of the protostars (typical age $\sim$ 10$^{5}$ yr), however, since luminosity is proportional to the mass (e.g. from  the theoretical isochrones of \citet[]{bre12}, we find that $\mstar \propto L^{4}$ for the stars in the mass range 1--10 \Ms~and 
an age of 10$^{5}$ yr), thus, we considered that any evidence of
luminosity segregation  is equivalent to mass-segregation. We used only $Herschel$ protostars to test the mass-segregation effect, as we have luminosity measurements of only these protostars. 
We acknowledge that in this simple mass-luminosity relation, we have ignored the role of accretion luminosity on the total luminosity of the protostars, but it is expected to be around 25\% for the class I sources \citep[e.g.][] {hill-white_2004}. 

We calculated $\Lambda_{MSR}$ starting at N (number of most massive stars) = 5 up to the number of protostars in our sample and calculated $<l_{random}>$ by picking 500
random sets of N stars.
Figure \ref{fig_msr} shows the $\Lambda_{MSR}$ for increasing values of N.  As can be seen, the 
8 most massive stars show the maximum  value of $\Lambda_{MSR}$ (i.e. $\sim$ 3.2 $\pm$ 0.5) , 
then $\Lambda_{MSR}$ progressively decreases and becomes flat beyond 
18 most massive stars. The larger $\sigma_{random}$ is expected for small N due to stochastic effects in choosing the small random sample \citep{ali2009}. 
Like for the analysis of the Q parameter, we also estimated the effect of 70 $\mu$m point source sample completeness on the mass-segregation and found that $\Lambda_{MSR}$ to be around 2.8, which is though on
the lower-side of the $\Lambda_{MSR}$ measured for the entire
cloud but within the error.
This analysis tells that the 8 most massive stars 
of \cloud~are likely 3 times closer to each other compared to the typical separation of 8 random stars in the region, suggesting that the mass-segregation effect is likely present in \cloud. These eight most massive stars (L $\geq$ 50 \lsun) of
\cloud~are located within  $\sim$ 9 pc from the adopted centre.
We discuss the likely cause of the observed mass-segregation in Section \ref{hub}. It is worth mentioning that, like here, mass-segregation of cores has been investigated in a few filamentary environments using ALMA observations involving a small number of cores. For example, \cite{plunk2018} reported that massive cores of
Serpens South are mass segregated with a median $\mathrm{\Lambda_{MSR}}$ 
of $\simeq$ 4. Similarly, \cite{sad2020} also 
find evidence of mass segregation in the filaments
of NGC 6334 with $\mathrm{\Lambda_{MSR}}$ value in the range $\simeq$ 2--3. 
However, we note that
within the \cloud~cloud area, the SFOG survey has identified 48 protostars, 31 of which have no counterparts in the 70 $\mu$m catalogue. These are likely the low-luminosity sources of the cloud beyond the sensitivity limit of the 70 $\mu$m image. 
Since the SFOG survey has  used data in the wavelength range 1--22 $\mu$m to identify these protostars, robust estimation of their bolometric luminosity is not possible. We thus refrain from using these sources in the MST analysis, but we acknowledge that the non-inclusion of these protostars and also any embedded low-luminosity protostars that are not detected in the SFOG survey may bias our results. Future high sensitivity multi-band long-wavelength observations are needed for a more precise estimation of the mass segregation.

\begin{figure}
\centering{
\includegraphics[width=8.5cm]{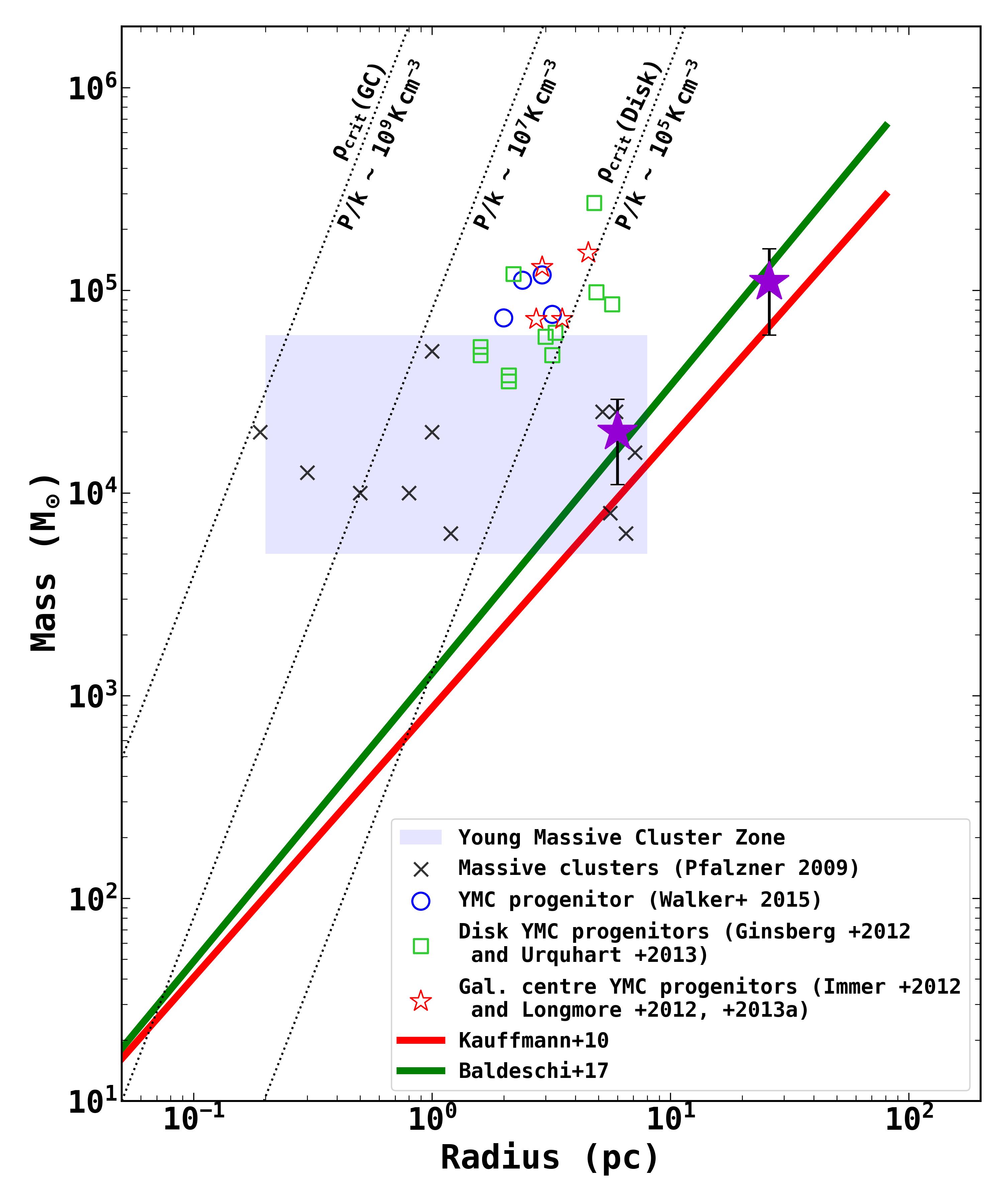}}
\caption{Mass-radius relation for massive star formation and YMCs in the Milky Way.  The red line is the threshold for massive star formation from
\citet{kauffmann_pillai2010}, while the green line is from  \citet{bald2017} threshold. The hatched rectangle shows the location of Galactic young massive clusters tabulated in \citet{pfal2009}. The green squares are YMC progenitor candidates in the disk \citep{Gins2012, Urq2013}, and the blue circles and red stars are the YMC progenitors in the Galactic Center from \citet{walk2015} and \citet{Immer2012, long2012, long2013}, respectively. Dotted lines show the predicted critical volume density thresholds for the GC, intermediate region, and the Galactic disk, assuming pressures of P/k $\sim$ 10$^9$, P/k $\sim$ 10$^7$, and P/k $\sim$ 10$^5$ K cm$^{−3}$,
respectively. The locations of \cloud~are shown in purple star symbols, corresponding to the mass measured at \ak = 0.2 mag and 0.8 mag, respectively. } 
\label{fig_hmt}
\end{figure}

\begin{figure*}
\centering{
\includegraphics[width=\textwidth]{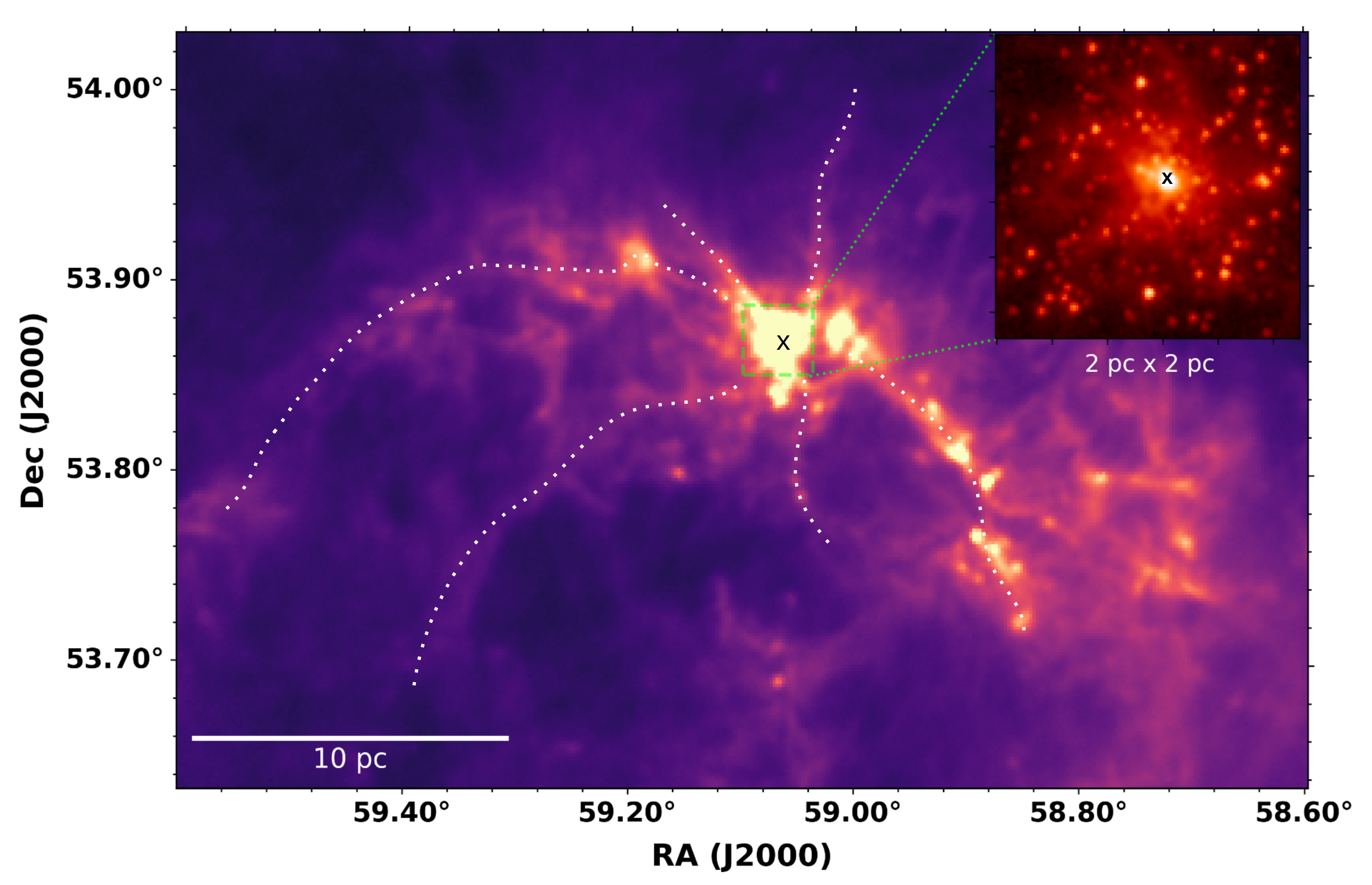}}
\caption{$Herschel$ 250 $\mu$m image of \cloud, revealing the filamentary structures in its central area. The inset image shows the zoomed-in view of the central region in $Spitzer$ 3.6 $\mu$m, which is taken from GLIMPSE360 survey \citep[]{whit_glimpse_2009}. It shows
the presence of an embedded cluster in the hub. The cross sign shows the position of the massive YSO.
}
\label{fig_250}
\end{figure*}

\section{Discussion}
\label{cfm}
The \cloud~cloud is massive and bound, yet  it is still speculative to say whether it will
form a high-mass cluster. Here, we consider a cluster with a total stellar mass $\geq$ 10$^4$ \Ms~as a massive cluster, following the classification scheme of \citet[][]{por10}. 



Simulations suggest that a high-mass cluster may form: (a) if the cloud collapses to form a centrally condensed  massive dense clump, which fragments
to form stars at a 
high efficiency \citep{ban15,ban18}. By doing so, the clump may produce a rich cluster in a short span of time before the stellar feedback commences  
and the process is called as ``monolithic'' or ``in-situ'' mode of cluster formation \citep[e.g.][]{ban15,walk2015}. In this scenario, the cluster forms at higher initial stellar densities, and then relaxes to its final state after it has expelled the gas.

(b) In the literature, many large-scale dynamical models involving the evolution of molecular clouds over an extended period of time have been proposed for making massive clusters. These include models such as global hierarchical collapse \citep[e.g.][] {vaz19}, gravitationally-driven gas inflows \citep[e.g.][] {smith2009}, and
conveyor-belt collapse \citep{long14,walk2016,barnes_2019,krum-mck2020}.
All these models have some similarities and differences  \citep[for details, see the review article by][]{vaz19}.
Broadly, these models point to the formation, evolution, and coalescence/convergences of substructures within a molecular cloud
into a more massive structure driven by global collapse, leading to the formation of  massive stars and  associated cluster. 
These models broadly suggest that, while the molecular cloud globally evolves, due to its  hierarchical nature, it also
simultaneously forms stars at local high-density structures (i.e. within the filaments or dense regions), as their free-fall times are shorter than the free-fall time of the global cloud. 
And as the evolution of the cloud proceeds, the cold matter in the extended environment and the protostars  formed within them can eventually be transported to the remote collapse centre, located at the cloud's centre of potential. This can occur via filaments, anchored by large-scale global collapse \citep[for details, see review articles by ][]{pin22, hac22}, where filaments 
act like conveyor-belts. 
In these scenarios, star formation would proceed over several crossing times leading to significant age spread in cluster members, the 
seeds of massive stars are expected to be located near the cloud's center of potential, and more younger stars are expected to form in the end. As a consequence, primordial mass segregation is expected, and also, the cloud's central potential is expected to have more young stars
compared to the stars in the extended part of the 
cloud \citep[e.g., see discussion in][]{vaz19}.

In the following, we discuss the possible scenarios of massive star and associated 
cluster formation in \cloud, and discuss our results in the context of the above
cluster formation theories.

\subsection {Observational Evidence of Massive Cluster Formation and Processes Involved in G148.24+00.41}

\subsubsection {Evidence of High-Mass Star and Associated Cluster Formation}
\label{hms}
Observations suggest that the majority of the massive stars form in clusters \citep{lada03} and also, the mass of the most massive star of a cluster is proportional to the total mass of the cluster \citep[e.g.][]{weid2010}. Therefore, the presence of young massive star(s) in a cloud is an indication of ongoing 
cluster formation. Another way of finding whether or not  a cloud would form a high-mass cluster is to look  for
mass versus radius diagram as it is suggested that to form a high-mass cluster, a reservoir of 
cold gas concentrated in a relatively small volume is likely required \citep[e.g.][]{bressert12,Gins2012,Urq2013}.


Based on column density maps derived from dust emission (MAMBO and Bolocam) and extinction (2MASS) data, \citet[][]{kauffmann_pillai2010} suggested  a  criterion for massive star formation. They argued that the clouds  known to be forming massive (M$_* \sim 10$ \Ms) stars have 
structural properties  described by $m(r) > 870$ \Ms $(r/\rm pc)^{1.33}$,
where $m(r)$ is the mass within radius $r$. Clouds below this criterion are unlikely to form massive stars. A similar conclusion is also given by \citet[][]{bald2017} for cold structures to form high-mass stars. 
In Figure \ref{fig_hmt}, we show these empirical relations along with the location of the \cloud~cloud corresponding to its total mass and dense gas mass.
 As can be seen, both the estimates of \cloud~lies nearly above these relations, suggesting that massive star formation is expected in \cloud. However, it is worth noting that  simulations suggest that the star formation efficiency, though highly dependent on the initial conditions, is usually low at the very initial stages of cloud evolution and accelerates after a few free-fall time  \citep{zamora12,lee2016, cald-chan2018, clark21}. In addition, models also suggest that compared to low-mass stars, the massive stars form last in a molecular cloud \citep{vaz09,vaz19} which is supported by
 some observations \citep[e.g.][]{fos14}, however, there
are also contrasting observational results suggesting that massive stars 
may form in the early phases of the molecular clouds \citep[e.g.][]{Zhang_2015}. 


All the above
models point to the fact that the non-detection of  high-mass stars in a massive cloud does not imply that it would not form high-mass stars. It may simply be due to the fact that the cloud is at the very early stages of its evolution and has not had enough time to form massive stars. Nonetheless, in the 
present work, 
the most luminous 70 $\mu$m point source of our sample corresponds to a probable massive YSO (MYSO; RA = 03:56:15.36, Dec = +53:52:13.10) listed in the MYSO sample of the Red MSX Source (RMS) survey \citep{lums13}.   \citet{coop13} confirms the YSO nature of the Red MSX Source using near-infrared spectroscopy observations. 
Scaling the luminosity of the Red MSX massive YSO tabulated in   \citet{coop13} to our opted distance, 
we find its luminosity to be $\sim$ 4200 $\lsun$ (= $7300 \times (3.4~ kpc/4.5~ kpc)^2$), which is two times of our 70 $\mu$m flux-based luminosity estimation (see Section \ref{lf}).
The dynamical age of the MYSO based on the extent and the velocity of the outflow lobes, traced with the CO (J = 3--2) transition, is suggestive of a very young
age, around $\sim$ 10$^5$ yr \citep{maud15}. No UC\hii region has been detected in the 5GHz continuum image of the RMS survey.
The  typical age of the UC\hii region is around $\sim$ 10$^5$ yr. All these results imply that the MYSO is in its 
early stages of evolution. 

Figure \ref{fig_hmt} also shows the location of young massive stellar clusters \citep[Mass $>$ 5 $\times$ 10$^3$ \Ms~and Age $<$ 5 Myr; ][] {pfal2009} by the shaded area. Also shown are the young massive cluster (YMC)  precursor clouds of the Galactic disk \citep{Gins2012, Urq2013} and Galactic Center \citep[GC;][]{Immer2012, long2012, long2013, walk2015}.  The YMC precursor clouds that have been identified at the Galactic Centre, are mostly quiescent, despite tens of thousands of solar masses of gas and dust  within only a few parsecs. 
 
It is believed that massive GC clouds are favourable places for YMC formation. It hosts the two most young massive (mass $\sim$ 10$^4$ \Ms) clusters in the Galaxy, the Arches and Quintuplet, which have formed in the GC recently, with ages of $\sim$ 3.5 and 4.8 Myr, respectively \citep[e.g.][]{walker2018}. As can be seen from Figure \ref{fig_hmt}, in terms of mass and compactness, compared to Galactic center clouds, the dense gas mass of \cloud~is lower by an order of magnitude while its radius is higher by a factor of 2--3. In \cloud, star formation is underway, as evident from the detection of YSOs of various classes, therefore, some of the gas has already been consumed in the process. Even then, comparing the current location of \cloud~with the location of the YMC progenitor clouds in the disk and GC, it appears that \cloud~may not form a YMC like the Arches cluster (mass $>$ 10$^4$ \Ms~and radius $\sim$ 0.5 pc). We compared the mass surface density profile of \cloud~within 2 pc from the hub center with other Galactic YMC precursor clouds discussed in \cite{walk2016}, and found that the  surface density profile of \cloud~is substantially below all of the Galactic centre clouds, and the extreme cluster forming regions in the disc. This again points to the fact that although \cloud~has a significant mass reservoir, but it is spread over a larger projected area, hence the lower surface mass density, and lower potential for forming a star cluster like the Arches.

The figure also shows the turbulent pressures for the different environments in our Galaxy \citep[for details see][]{long14}. Assuming that \cloud~is pressure confined by the turbulent pressure of the Galactic disk, to become unbound, the internal pressure of the cloud has to be of the order of 10$^5$ K cm$^{-3}$. 
The present dynamical status of \cloud~suggests that it is gravitationally bound. 
In the following, we explore what kind of cluster \cloud~may form. 

\subsubsection{Likely Age and Mass of the Total Embedded Stellar Population}
\label{age}
To understand the stellar content of the hidden embedded population, we again make use of the SFOG survey YSO catalogue. In the field of \cloud, SFOG survey has identified 175 YSOs, out of which 48 are class 0/I, 120 are class II, and 7 are class III sources. We matched and combined the SFOG YSO catalogue with the protostars identified in this work, which resulted in a total of 187 YSOs, out of which 70 are found to be protostars. 

\begin{figure}
    \centering
    \includegraphics[width=8cm]{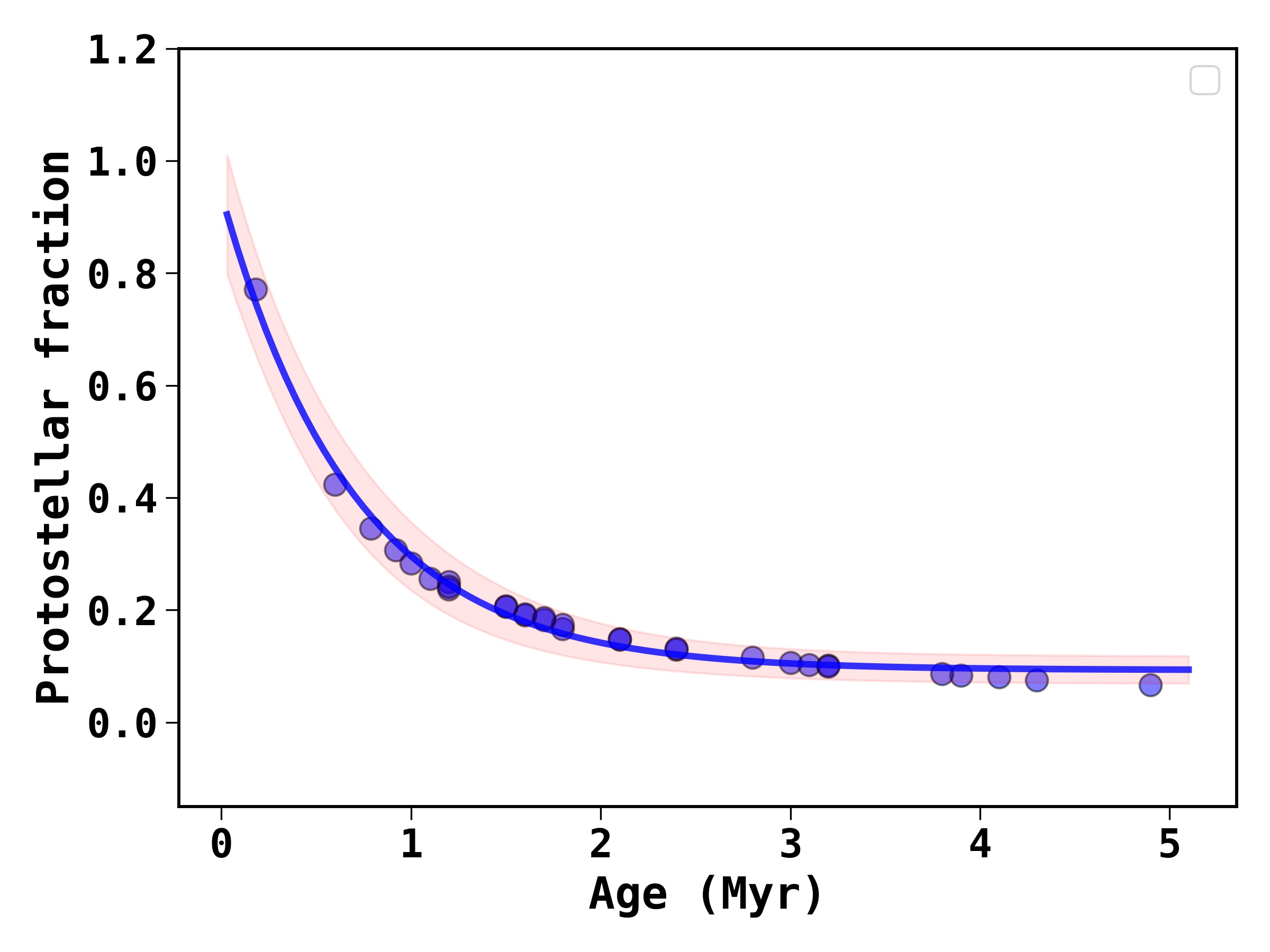}
    \caption{Age of stellar sample vs. fraction of protostars. The blue line shows the best-fit exponential decay curve (see text for more details).}
    \label{fig_pf}
\end{figure}

Young stellar objects take different amounts of time to progress through the various evolutionary stages. Protostars (Class 0, Class I, and flat-spectrum sources) represent an earlier stage of young stellar object's evolution than the class II and class III sources, thus, the ratio of 
protostars  to the total number of YSOs  (or class II sources) are often used to derive 
relative ages of the star-forming regions \citep[e.g.][]{jor06,gut08,mye12}. 
For deriving an approximate age of \cloud, we compare its protostellar fraction with that of 
the well-known star-forming regions.
Figure \ref{fig_pf} shows the protostellar fraction (i.e. ratio of protostars to  
the total number of protostars plus class II YSOs) vs. age of the 23 
star-forming regions tabulated in \citet{mye12}. In this figure,  class III sources are not
considered in the total number of YSOs following \citet{mye12}, as the authors did not consider 
class III sources in their analysis, arguing that they are incomplete.  
As can be seen from Figure \ref{fig_pf}, the protostellar fraction declines with age.  Assuming that the protostellar fraction decays exponentially with the age, like the disk fraction in young clusters \citep{Rib2014}, we fitted the observed protostellar fraction as a function of time using an exponential law of the form: 
\noindent
$f_{pro}$ = $A\, exp\left(\frac{-t}{\tau}\right) + C$
\noindent, where $t$ is the age of the sample (in Myr), $A$ is the initial protostellar fraction, $\tau$ is the characteristic timescale of decay in protostellar fraction (in Myr), and $C$ is a constant level. 
The best-fitted value of $A$, $\tau$, and $C$ are 0.847 $\pm$ 0.022, 0.700 $\pm$ 0.024, and 0.093 $\pm$ 0.005, respectively. The derived relation is an indication of the fact that 
the protostar's life-time is around $\sim$ 0.7 Myr. For \cloud, we estimated the protostellar fraction to be $\sim$ 37\% with an admittedly high uncertainty, which is difficult to quantify considering the likely completeness limits of various bands used in the SFOG survey for
identifying the YSOs and also the sensitivity limits of these bands in detecting disk-bearing stars in the cloud due to its distant nature.
Nonetheless, taking the observed protostellar fraction a face value and using the above-derived relation, we make a crude assessment of the age of \cloud~to be roughly around 1 Myr, which we used in this work.

Measuring the mass of individual embedded YSOs is an extremely challenging task in young star-forming clouds due to the presence of variable extinction within the cloud and also
the presence of infrared excess in YSOs. Nonetheless, to get
a rough census of the total stellar mass that might be embedded in the cloud, we first estimated the typical detection limit of our YSO sample.  We do so, by 
searching the counterparts of the YSOs in UKIDSS near-infrared bands, adopting  1 Myr as their age and assuming a minimum foreground \av of 5 mag (which corresponds to the outer column density
boundary of the central area, within which the majority of the YSOs are concentrated) is present in the direction of the YSOs. We find that the typical detection limit is around 0.9 \Ms.  Here, we used 1Myr theoretical isochrone from
\citet{bre12} and corrected the isochrone for the distance and extinction to compare with the observed
near-infrared magnitudes of the YSOs. 

The luminosity of the most massive YSO is around $\sim$
1900 \lsun,  which corresponds to a star of 8 \Ms~\citep[see Table 1 of][]{Mott_2011}. Considering that there are 187 point sources embedded in the cloud between 0.9 and 8 \Ms~and using the functional form of Kroupa mass-function \citep{kroupa_2001}, we estimated the total stellar mass to be $\sim$
500 \Ms~and extrapolating down to 0.1 \Ms, we find the total mass to be $\sim$ 1000 \Ms. Thus, we expect an embedded population of total stellar mass around 1000 \Ms~to be present in the cloud.
We note that applying a higher foreground extinction would give even a higher mass detection limit for YSOs and, thus, a higher total stellar mass. 

\subsubsection{Hub Filamentary System and Its Implication on Cluster Formation}
\label{hub}

Figure  \ref{fig_250} shows the central area of
the cloud at 250 $\mu$m. In the image, several
large-scale filamentary structures (length $\sim$ 5--20 pc) were found to be 
apparent. These structures are sketched in the figure by the dotted curves and
meant only to indicate the possible existence of large-scale filament-like structures. 
These structures were identified by connecting nearly continuous dust emission structures present in the cloud.  A thorough identification of the filaments is beyond the scope of the present work.
Future high-dense gas tracer molecular data would be highly valuable for identifying the velocity coherent structures, thus the filaments in the cloud and their properties \citep[e.g.][]{tre19}. 
However, from the present generic sketch, one can see that the central
dense location is located at the nexus of six filamentary structures. We  find
that this central area is host to an embedded cluster, seen in the near-infrared images. The inset image of Figure \ref{fig_250} shows the cluster in the $Spitzer$ 3.6 $\mu$m band, and as can be seen, it contains a rich number of near-infrared point sources with massive YSO at its very center.
Altogether,  the whole morphology of Figure \ref{fig_250} is  consistent with the picture of a hub filamentary system put forward by \citet{mye09}, where several fan-like filaments are expected to intersect, merge
and fuel the clump located at their geometric centre \citep[e.g. see also discussion in][]{kumar22}.

As discussed in Section \ref{ms}, we have observed the evidence of mass-segregation in the cloud for luminous sources with luminosity $>$ 50 \lsun~within 9 pc from the central hub. 
This observed mass-segregation could be of primordial or dynamical origin. In case of star clusters, the dynamical origin is primarily driven by the interaction among the stars. In the present case,
the total mass of the embedded population is around
$\sim$ 1000 \Ms, which is $\sim$3\% of the 
total gas mass ($\nht > 5 \times 10^{21} \cms$)
enclosing these YSOs. This is suggestive of the
fact that the gravitational potential 
in the central area of the cloud
is dominated by the gas than the stars; thus, dynamical interaction among the stars might not be so effective at this stage of the cloud's evolution for global mass-segregation to happen. The cold molecular  gas and dust are usually thought to impede the process of dynamical interaction.\\

Next, we look for whether the observed mass-segregation is driven by filamentary flows.  Because filaments are often associated with longitudinal flows \citep[e.g.][]{per13,dut18,rya18}, heading toward the bottom of the potential well
of the system \citep[e.g.][]{tre19}. We calculated the flow crossing
time as: $t_{\rm cr}$ =$ \frac{R}{v_{\rm {inf}}}$, where v$\mathrm{_{inf}}$ is
the flow velocity and R is the distance travelled by gas flow.
In the absence of v$\mathrm{_{inf}}$ information for \cloud, we used the typical value of 
v$\mathrm{_{inf}}$  in the range 1--1.5 \kms pc$^{-1}$, observed in the large-scale filaments that
are radially attached to the massive
star-forming hubs \citep[e.g.][]{tre19,mon19}, to calculate the flow travel distance. Doing so, we estimated that in 0.1--1 Myr of time (i.e. the likely age range of the region; discussed in Sections \ref{hms} and \ref{age}), the flow would
travel a distance in the range of 0.15 to 1.5 pc.  If this flow carries massive prestellar cores or massive protostars along with it, then one would expect that the effect of mass-segregation within 1.5 pc from the centre of the cloud's potential may be of flow origin. However, we want to stress that it is very unlikely that the massive prestellar core or protostars would flow along the filaments with the same velocity as gas particles may do. Thus, the aforementioned estimated travel 
distance would be an upper limit for the protostars. 

Since the mass-segregation scale (9 pc) for the massive stars is larger than the flow travel distance (0.15--1.5 pc) for the adopted age range of the system, we thus hypothesized
that the evidence of global mass-segregation observed in \cloud, if confirmed (see possible biases in Section \ref{ms}), may suggest towards its primordial origin. Deeper photometric observations along with the velocity measurements of the gas and protostars would shed more light on this issue.

\subsubsection {Prospects of Cluster Formation Processes in \cloud}
\label{dmcf}

Simulations suggest that the density profile reflects the physical processes influencing the evolution of a cloud. The overall density profile, $\rho \propto r^{-1.5}$,  obtained for \cloud~is a signature of
a self-gravitating turbulent cloud. This is also revealed by the distribution of the protostars. We obtained 
Q-value around 0.66 from the distribution of protostars, suggesting that the central region is 
moderately fractal with a fractal dimension equivalent to 2.2.
This fractal structure could be a consequence of both gravity and turbulence. For example, \cite{dobb05} simulated a turbulent clump of density profile,  $\rho \sim r^{-1.5}$ and found that the clump is able to fragment  into hundreds of cores that are tied with filamentary structures. Q-value $>$ 0.9 generally represents a steeper density profile of
exponent $\beta$ $>$ 2. Individual clumps
may have a steeper density
profile, but the central area as a whole is fractal.
In other words, the central area of the cloud is different from the profile that one would expect for a cloud to form a single compact cluster via monolithic collapse. The distribution  of protostars  across the length of the dense gas over a range of 
densities (see Section \ref{lf}) also disfavours a monolithic mode of cluster formation 
in \cloud. \citet{walk2015,walk2016} compared the profile of gas density distribution of the YMC precursor clouds with the stellar density distribution of the YMCs. They found that the density profile of the former
is flatter compared to the latter, which led them to suggest that the YMC precursors are not consistent with the monolithic 
formation scenario of star clusters. Doing a similar analysis, we found that
the gas surface density profile of the hub area of \cloud~is flatter than the stellar distribution of YMCs, supporting the above notion that the present cloud is not centrally concentrated enough to form a typical massive cluster in-situ given the present-day mass distribution.


\begin{figure*}
\centering{
\includegraphics[width=14cm]{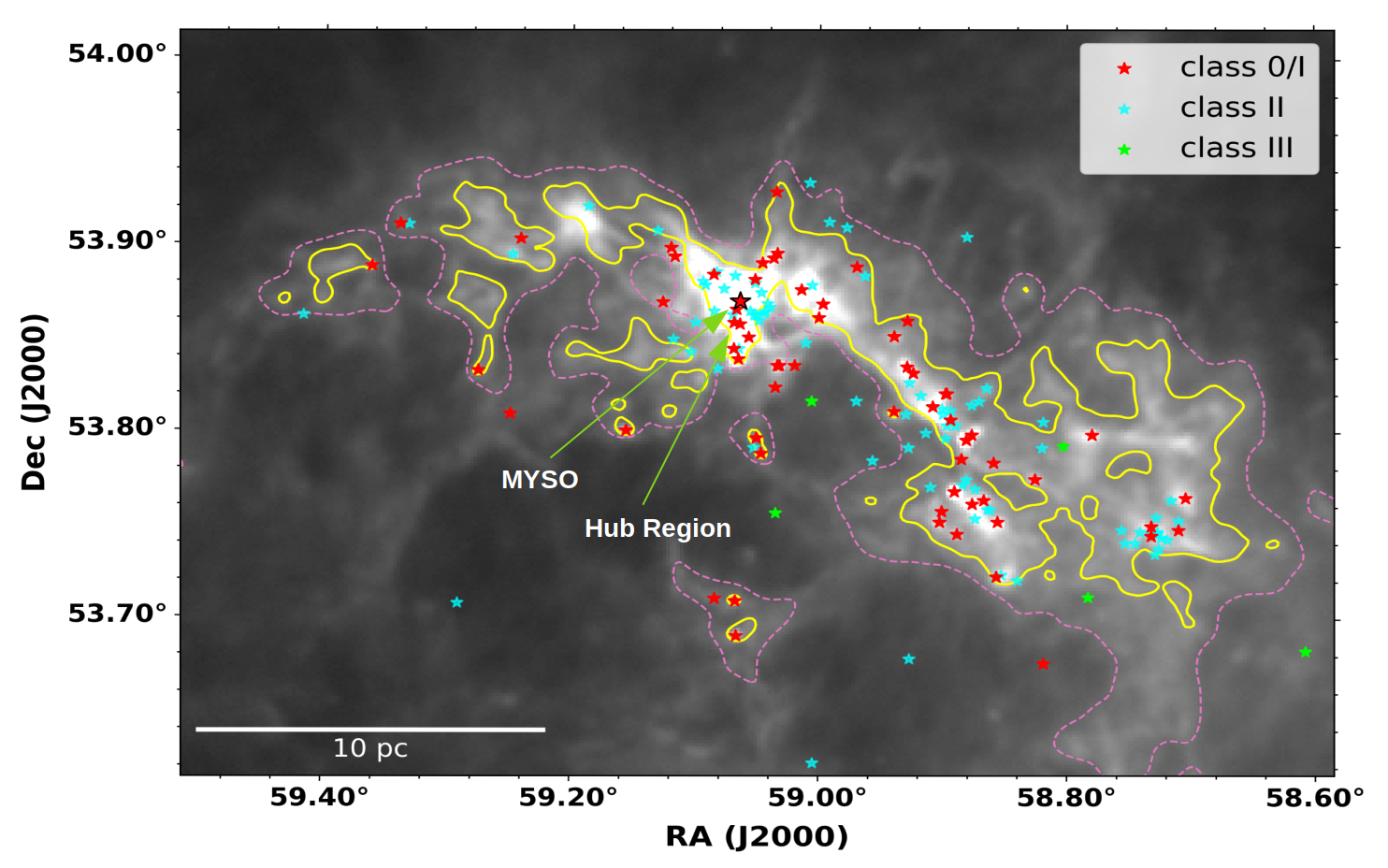}}
\caption{The distribution of YSOs from $Herschel$ 70 micron point source catalog \citep{hers2020} and SMOG catalog \citep{wins2020} on the $Herschel$ 250 micron image. The dotted pink color contour shows the column density at $5.0 \times 10^{21}$ cm$^{−2}$ and the yellow solid color contour shows the dense gas column density at $6.7 \times 10^{21}$ cm$^{−2}$ (\ak $\sim$ 0.8 mag). Protostars, class II, and class III YSOs are marked by red, cyan, and green star symbols, respectively. }
\label{fig_dist}
\end{figure*}

\begin{figure}
\centering{
\includegraphics[width=8.5cm]{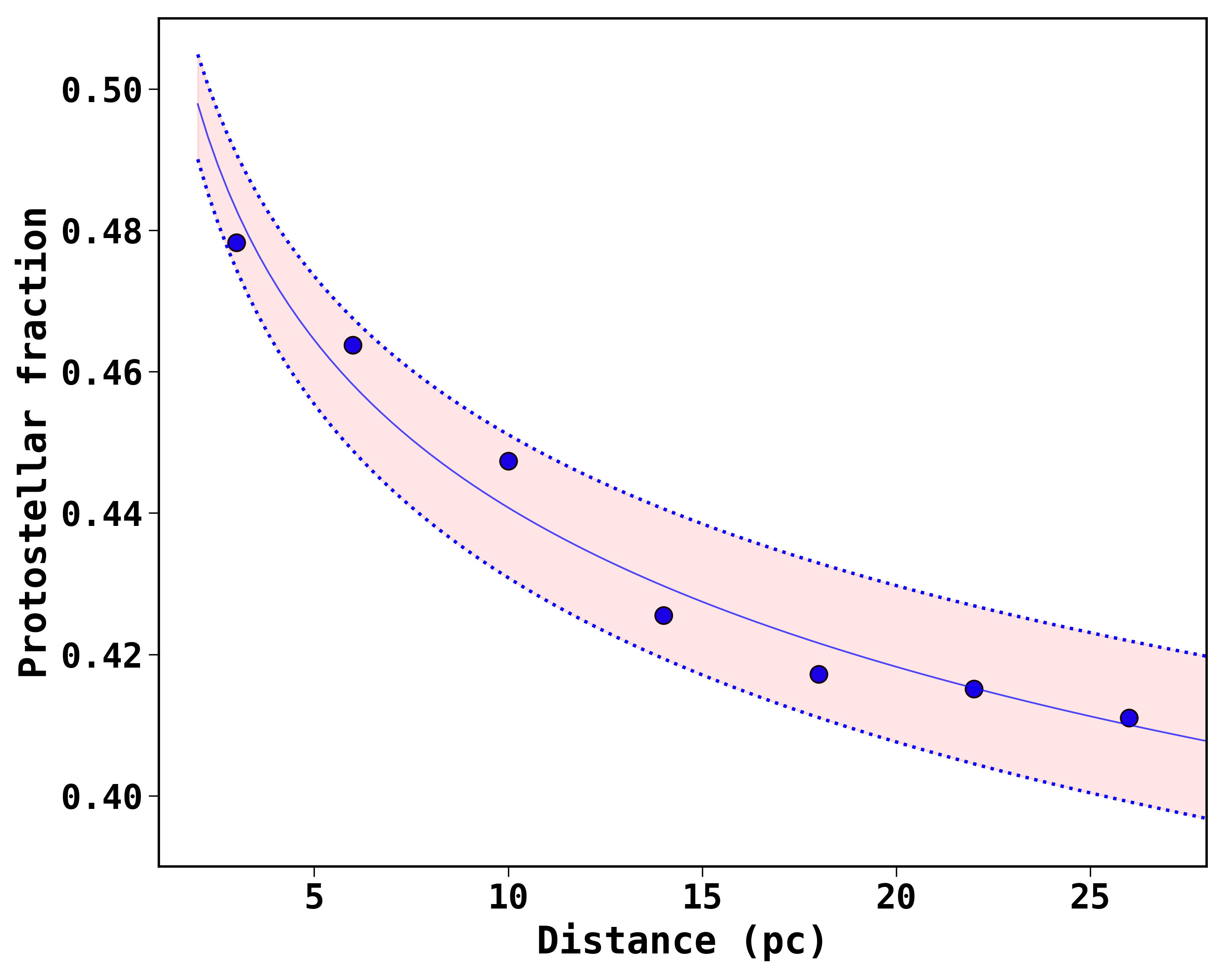}}
\caption{Radial distribution of protostellar fraction from the hub location. The blue solid line represents
to a power-law profile of index $\sim$ $-$0.08, while the 
shaded area represents the 1$\sigma$ uncertainty associated to the power-law fit.}
\label{fig_pro}
\end{figure}

In Figure \ref{fig_250}, we show that the \cloud~cloud hosts a hub-filamentary system, where cluster formation is happening at the hub of the filaments. The presence of hub-filament systems has also been advocated in flow-driven simulations including global collapse \citep{smith2009,gom14,vaz19}. From the evidence of the hub filamentary system, density profile with a power-law index of $-$1.5, and low Q-value at the central area, it appears that the whole cloud may be self-gravitating globally. However, at smaller scales, the star formation is occurring in dense structures such as filaments and hubs that are immersed within this large-scale self-gravitating cloud. Moreover, though protostars have formed over a range of densities, the high-luminosity sources (or the high-mass sources) are located around the densest locations of the cloud. We find evidence
of mass-segregation in the \cloud~cloud. The low Q-value and the fact that the flow crossing scale is lesser than the mass-segregation scale suggest that mass segregation is likely primordial. In addition, the massive star, which is still at a very young age, of the order of 10$^5$ yr, is found to be located in the central area
of the cloud, while the young class II sources whose age lies in the range $\sim$ 1--2 Myr have also been observed in the cloud \citep{wins2020}. 

Figure \ref{fig_dist} shows the distribution of the
YSOs on the $Herschel$ 250 $\mu$m image. As can be seen, most of the YSOs are located  near the hub or associated filaments. We want to stress that the detection limit of our identified YSOs is around 1 \Ms.  Thus, there may be more faint low-mass 
YSOs distributed in the extended part or diffuse filaments of the cloud and are not identified here. Also, due to the crowding of stellar sources and bright infrared background, the true YSO number identified inside the hub using {\it Spitzer} images may be an underestimation.
Nonetheless, from the present sample, we calculated the protostellar fraction
as a function of distance from the central hub, and is shown in Figure \ref{fig_pro}.  As can be seen, the 
plot 
signifies that younger sources show
the tendency of being located  closer to the cloud centre relative to the class II YSOs.
All the above evidence point to the flow-driven modes of cluster formation that are discussed in Section \ref{cfm}. So, we hypothesize that, if the cloud will ultimately form a high-mass cluster, it has to go through global hierarchical convergence and merger of its both gaseous and stellar content as advocated in conveyor-belt type models \citep[e.g.][]{long14,walk2016,vaz19,barnes_2019}. The latter can even occur after no gas is left in the system if the stellar sources are part of a common potential \citep[e.g.][]{how18,sills18,karam22}.

\subsection {Predictions from Models of Hierarchical Star Cluster Assembly and Merger}
\label{ps}

Assuming that the cloud will form a high-mass cluster through dynamical processes over an  extended period of time (over a few Myr), involving global hierarchical collapse and merger of stars and subgroups, then it is tempting to speculate that what kind of cluster it may form.

\citet{gav17} studied the early (up to 2 Myr) dynamical evolution of a turbulent cloud of mass $2.5 \times {10}^{4}$ \Ms, radius $\simeq$ 5 pc, and 3D velocity dispersion $\simeq$ 2.5 \kms. They found
that as the cloud collapses, it forms stars in filaments and extended part of the cloud at a slow rate, but a rich high-mass star cluster emerges from the cloud at the end of the simulation that has some features similar to the massive cluster NGC 6303. In terms of mass, radius, and velocity dispersion, the properties of the simulated cloud are nearly the same as the dense gas properties of \cloud (see Section \ref{dgf}), implying that \cloud~has the potential to build a rich  cluster. 

From an observational point of view, the emergence of a massive star cluster also seems to be feasible for \cloud, because
its embedded stellar mass is $\sim$
1000 \Ms, while it still has a high reservoir of bound gas to make more stars. 
If we go by the hypothesis that it is the dense gas that contributes more to star formation, then using the relation between star formation rate and dense gas mass of \citet{lada12}:
\begin{equation}
\label{sfr}
SFR = 4.6 \times 10^{-8} f_\mathrm{den} M_{tot}(M_\odot)\  \ M_\odot\, yr^{-1},
\end{equation}
one could make a rough assessment of the total stellar mass that may emerge from the cloud. The $M_{tot}$ and $f_{den}$ are the total mass and dense gas fraction of the cloud, respectively. By taking dense gas fraction $\sim 18\%$, total gas mass $\sim$ $1.1 \times 10^5$ \Ms~(see Section \ref{dgf}), and assuming star formation would proceed at a constant rate for another 1 to 2 Myr, we find that a stellar system of total mass in the range 1000--2000 \Ms~may emerge from \cloud.
This prediction is also in line with the recent simulation results of \citet{how18}. \citet{how18} follow the evolution of massive GMCs (mass in the range 10$^4$--10$^7$ \Ms) with feedback on and off. They found that the star clusters 
emerge from the cloud via a combination of filamentary gas accretion and mergers of less massive clusters, and found a clear relation between the maximum cluster (M$_{\rm{max}}$) mass and the mass of the host cloud (M$_{\rm {GMC}}$). Observations also support this prediction \citep[e.g.][]{he22}. Following the prediction of \citet{how18} for feedback on and considering
the dense gas mass only as the cloud mass, we find that the dense gas reservoir has the ability to form a cluster of total stellar mass  $\sim$ 2000 \Ms. 

Combining the embedded stellar mass and the expected 
stellar mass from the present dense
gas reservoir, one would expect a total stellar mass in the range 2000--3000 \Ms~to emerge from this cloud. 
We note that this is the case, without accreting any additional gas from the extended low-density reservoir beyond the effective radius 
of the dense gas (i.e. $\sim$ 6 pc). However, considering the fact that molecular clouds are highly dynamical, if the cluster accretes cold gas from the extended reservoir, then the total stellar mass is likely an underestimation. 

We want to stress that simulations of cluster-forming clouds have shown that  molecular clouds tend to have some degree of fractal structures at their early stages of evolution, as found in \cloud. But the degree of fractality slowly reduces as the evolution proceeds because gravitational collapse together with stellar dynamical interactions among the stars and the subgroups progressively erase the initial conditions of
the cloud and build up a dense and spherical star cluster \citep[e.g.][]{mas10,gav17,how18}. If this happens for \cloud~in future, where most of the stellar sources segregate to a cluster at the bottom of the potential well, we may expect a centrally condensed massive cluster with Q-value $>$ 1,  otherwise, it may evolve to a massive association 
of stars or groups. 

Although, these predictions suggest that the cloud has the potential to form a rich cluster in the range 2000--3000 \Ms, yet further studies of the cloud concerning its gas properties and kinematics are necessary for investigating whether the filaments that appear in the dust continuum images are indeed converging and funneling the cold matter to the central potential of the cloud. Thus, would facilitate the formation and emergence of a
dense cluster like the ones predicted in the above simulations.

\section{Summary and Conclusions}
\label{s28_conc}
We have studied the global properties and cluster formation potency of the \cloud~cloud using dust continuum and dust extinction measurements. From both the dust continuum and dust extinction map, we found that the cloud is massive (M $\sim$ 10$^5$ \Ms) and has dust temperature $\sim$ 14.5 K, radius $\sim$ 26 pc, and surface mass density $\sim$ 52 \Ms~pc$^{-2}$. It follows the power-law density profile with index $\sim$ $-$1.5 and has a virial mass almost half of the total cloud mass, which shows that it is likely bound. By comparing \cloud~with other galactic molecular clouds, studying its clustering structure, identifying its protostellar content, and inspecting different cluster formation scenarios, we found that:

\begin{enumerate}

\item \cloud~shows high gas mass content, which is  comparable to GMCs like Orion-A, Orion-B, and California, and higher than the other nearby molecular clouds. From the dust continuum map, its dense gas fraction was found to be $\sim$ 18\%, which is slightly smaller than Orion-A and RCrA molecular clouds, but relatively higher than the other nearby MCs. The mass and effective radius of the cloud follows the Larson's relation, which is in agreement with the other nearby MCs.

\item We identified 40 protostars based on $\it{Herschel}$ 70 $\mu$m data.
By including the SFOG survey, the total number of protostars reaches to 70. Using MST analysis over these protostars, we found that the Q value is $\sim$ 0.66, which shows that the clustering structure is moderately fractal or hierarchical.

\item The spatial distribution of protostars shows that most of them are located in the central area of the cloud above \nht~$>$ 5 $\times$ 10$^{21}$ cm$^{-2}$. And the luminosity distribution shows that the high luminosity sources are relatively closer to the cloud centre and located in high surface density regions compared to low luminosity sources. This indicates the signature of mass segregation in the cloud, and we obtained the degree of mass segregation around 3.2 using MST technique.

\item After including the YSOs identified in the SFOG survey using GLIMPSE360 field, we estimate 
the likely total mass of the stellar population embedded in the cloud to be around 1000 \Ms, and by including the further star formation from dense gas only, we speculate that \cloud~has the potential to form a cluster in the mass range of 2000--3000 \Ms.

\item The cloud possesses a hub filamentary system, and a young cluster is seen in NIR at the hub location, along with the MYSO of $L=1900\, \lsun$. Along with the mass segregation, the younger sources are  closer to the cloud's centre of potential than the older ones. These findings, along with the evidence like low Q value and density profile, point towards the possibility of conveyor belt type 
mode of cluster formation in the \cloud~cloud. 

\end{enumerate}
\section*{ACKNOWLEDGEMENT}

The research work at Physical Research Laboratory is funded by the Department of Space, Government of India. This work makes use of data obtained from the UKIRT Infrared Deep Sky Survey, obtained using the wide field camera on the United Kingdom Infrared Telescope on Mauna Kea. We thank Stefan Meingast for the discussion on using PNICER algorithm to generate the extinction maps. We acknowledge the Herschel Hi-GAL survey team and ViaLactea project funded by EU carried out at Cardiff University. 
DKO acknowledges the support of the Department of Atomic Energy, Government of India, under Project Identification No. RTI 4002. We thank Eugenio Schisano for providing the Herschel Hi-Gal column density and dust temperature map of the \cloud~cloud region. We are thankful to S.N. Longmore for sharing the data of massive molecular clouds, pressure lines, and other critical parameters used in their work to compare them with \cloud. This research made use of the data from the Milky Way Imaging Scroll Painting (MWISP) project, which is a 
northern galactic plane CO survey with the PMO-13.7m telescope. We are grateful to all the members of the MWISP working group, particularly the staff members at PMO-13.7m telescope, for their long-term support. MWISP was sponsored by National Key R\&D Program of China with grant 2017YFA0402701 and CAS Key Research Program of Frontier Sciences with grant QYZDJ-SSW-SLH047.

\section*{Data Availability}

We used the NIR, FIR, and CO molecular data from UKIDSS, {\it Herschel}, and PMO, respectively, in this work. The UKIDSS and {\it Herschel} data are publicly available. The PMO data can be shared by the PMO database on reasonable request. We have also used the GLIMPSE360 data of the SFOG survey, which is publicly available on SFOG Dataverse.    



\bibliographystyle{mnras}



\appendix


\bsp	
\label{lastpage}

\end{document}